\documentclass[apj]{emulateapj}

\usepackage{apjfonts}
\usepackage{natbib,graphicx,epsfig,rotating,color,verbatim}

\def\fermi{{\it Fermi}}

\newcommand{\lsi}{LS~I~+61$^{\circ}$303}
\newcommand{\ls}{LS\,5039}

\begin{document}
\title{Long-term monitoring of the high-energy $\gamma$-ray emission from \lsi\ and \ls}

%

\author{D. Hadasch\altaffilmark{1}, D. F. Torres\altaffilmark{1,2}, T. Tanaka\altaffilmark{3}, R. H. D. Corbet\altaffilmark{4,5}, A. B. Hill\altaffilmark{3,6}, R. Dubois\altaffilmark{3}, G. Dubus\altaffilmark{7}, T. Glanzman\altaffilmark{3}, S. Corbel\altaffilmark{8}, J. Li\altaffilmark{9}, Y. P. Chen\altaffilmark{9}, S. Zhang\altaffilmark{9}, G. A. Caliandro\altaffilmark{1}, M. Kerr\altaffilmark{3}, J. L. Richards\altaffilmark{10}, W. Max-Moerbeck\altaffilmark{10}, A. Readhead\altaffilmark{10}, G. Pooley\altaffilmark{11}}
\altaffiltext{1}{Institut de Ci\`encies de l'Espai (IEEC-CSIC),
              Campus UAB,  Torre C5, 2a planta,
              08193 Barcelona, Spain}
\altaffiltext{2}{Instituci\'o Catalana de Recerca i Estudis Avan\c{c}ats (ICREA)}
\altaffiltext{3}{W. W. Hansen Experimental Physics Laboratory, Kavli Institute for Particle Astrophysics and Cosmology, Department of Physics and SLAC National Accelerator Laboratory, Stanford University, Stanford, CA 94305, USA}
\altaffiltext{4}{NASA Goddard Space Flight Center, Greenbelt, MD 20771, USA}
\altaffiltext{5}{Center for Space Science and Technology, University of Maryland Baltimore County, Baltimore, MD 21250, USA}
\altaffiltext{6}{School of Physics and Astronomy, University of Southampton, Highfield, Southampton, SO17 1BJ, UK}
\altaffiltext{7}{UJF-Grenoble 1/CNRS-INSU, Institut de Plan\'etologie et d'Astrophysique de Grenoble, UMR 5274, 38041 Grenoble, France}
\altaffiltext{8}{Universit\'e Paris 7 Denis Diderot and Service d'Astrophysique, UMR AIM, CEA Saclay, F-91191 Gif sur Yvette, France}
\altaffiltext{9}{Key Laboratory of Particle Astrophysics, Institute of High Energy Physics, Chinese Academy of Science, Beijing 100049, China}
\altaffiltext{10}{Cahill Center for Astronomy and Astrophysics, California Institute of Technology, Pasadena, CA 91125, USA}
\altaffiltext{11}{Cavendish Laboratory, Cambridge CB3 0HE, UK}

\begin{abstract}

The \fermi\ Large Area Telescope (LAT) reported the first definitive GeV detections
of the binaries \lsi\ and \ls\ in the first year after its launch
in June, 2008.
These detections were unambiguous as a consequence of the reduced positional uncertainty and 
the detection of modulated $\gamma$-ray emission on the corresponding orbital periods. 
An analysis of new data from the LAT, comprising 30 months of observations,
identifies a change in the $\gamma$-ray behavior of \lsi. An increase in flux
is detected in March 2009 and a steady decline in the orbital flux modulation
is observed. Significant emission up to 30\,GeV is detected by the LAT; prior
datasets led to upper limits only. Contemporaneous TeV observations no longer
detected the source, or found it -in one orbit- close to periastron, far from
the phases at which the source previously appeared at TeV energies.
The detailed numerical simulations and models that exist within the literature do not predict or explain many of these features now observed at GeV and TeV energies.  New ideas and models are needed to fully explain and understand this behavior.
A detailed phase-resolved analysis of the spectral characterization of \lsi\
in the GeV regime ascribes a power law with an exponential cutoff
spectrum along each analyzed portion of the system's orbit. 
The on-source exposure of \ls\ is also substantially increased with respect to our prior publication. In this case, whereas the general $\gamma$-ray properties remain consistent, the increased statistics of the current dataset allows for a deeper investigation of its orbital and spectral evolution.

\end{abstract}



\section{Introduction}

To date there are only a handful of X-ray binaries that have been detected at high (HE;
0.1--100\, GeV) or very high-energies (VHE; $>$100 GeV): \lsi\ \citep{2006Sci...312.1771A, 2008ApJ...679.1427A, 2009ApJ...701L.123A}, \ls\ \citep{2005Sci...309..746A,2009ApJ...706L..56A}, PSR
B1259$-$63 \citep{2005A&A...442....1A,2011ApJ...736L..11A}, 
Cyg X$-$3 \citep{2009Sci...326.1512F}, Cyg X$-$1 \citep{2007ApJ...665L..51A,2010ApJ...712L..10S}.
Recently, two new binaries were found: 1FGL J1018.6$-$5856, with a period of 16.6 days found in the GeV regime \citep{The Fermi Symposium 2011} and HESS J0632+057 \citep{2011ATel.3152....1F,2011ATel.3153....1O,2011ATel.3161....1M}, for which a period of $\sim$320 days was detected in X-rays (Bongiorno et al. 2011).
Of these sources only  \lsi, \ls\ and PSR B1259$-$63 share the property of being binaries detected at both GeV and TeV energies. 
The other systems have been unambiguously detected only in one band, either at GeV or at TeV, see, e.g., the case of Cyg X$-$3 in \citet{2010ApJ...721..843A}.  
In the case of Cyg X$-$1, with the hint of TeV detection itself being at the level of 4 standard deviations (4$\sigma$), claims of detection at GeV energies by the Astrorivelatore Gamma a Immagini Leggero (AGILE) remain uncertain with concurrent \fermi\ Large Area Telescope (LAT) observations \citep{2010arXiv1008.4762H}. It is yet uncertain whether these spectral energy distribution (SED) differences reflect an underlying distinct nature, or are just a variability signature in different bands.

The nature of the binary compact object in \lsi, \ls, HESS J0632+057 and 1FGL J1018.6$-$5856 is as yet undetermined \citep{2010arXiv1008.4762H}.  Both neutron star (e.g. PSR B1259$-$63) and probable black hole (e.g. Cyg X$-$3) binary systems have been detected at GeV energies and so both types of compact object are viable in the undetermined systems.  Recently the Burst Alert Telescope (BAT) onboard Swift reported a magnetar-like event which may have emanated from \lsi\ \citep{2008GCN..8215....1B,Torres2011}.  If true this would be the first magnetar found in a binary system.

The early LAT reports of GeV emission from \ls\ and \lsi\ were based upon 6--9 months of survey observations \citep{2009ApJ...701L.123A,2009ApJ...706L..56A}. Both sources were detected at high significance and were unambiguously identified with the binaries by their flux modulation at the corresponding orbital periods,
26.4960 days for \lsi\ \citep{2002ApJ...575..427G} and 3.90603 days for \ls\ \citep{2005MNRAS.364..899C}.
The modulation patterns were roughly consistent with expectations from inverse Compton scattering plus $\gamma$-$\gamma$ absorption models, and were anti-correlated in phase with pre-existing TeV measurements \citep[e.g.,][]{2009ApJ...693..303A, 2006A&A...460..743A}. The anti-correlation of GeV--TeV fluxes is in fact a generic feature of these models, where the GeV emission is enhanced (reduced) when the highly relativistic electrons seen by the observer encounter the seed photons head-on (rear-on); \citep[e.g., see][]{2007A&A...464..259B,2007ApJ...671L.145S,2008A&A...477..691D,2008MNRAS.383..467K}. \fermi-LAT measurements provided a confirmation of these predictions.

The spectra of both sources were best modeled with exponential cutoffs in their high-energy spectra, at least along part of the orbit. Specifically, an exponential cutoff was statistically a better fit to the SED compared with a pure power law at phases surrounding the superior conjunction (SUPC) of \ls and in the orbitally averaged spectrum of \lsi. Statistical limitations of the data prevented the determination or the ruling out of an exponential cutoff in any part of the orbit of \lsi\ or in the inferior conjunction (INFC) of \ls. The spectral energy distributions with the exponential cutoffs that were reported  were reminiscent of the many pulsars the LAT has detected since launch \citep{2009ApJ...706L..56A}, although this was far from a proof of their pulsar nature. To date no pulsations have been found at GeV energies, or at any other wavelengths, despite deep dedicated searches (see e.g., Rea et al. 2010, 2011).

Since \fermi\ was launched, both the Major Atmospheric Gamma-ray Imaging Cherenkov Telescopes (MAGIC) and the Very Energetic Radiation Imaging Telescope Array 
(VERITAS) have performed observations of \lsi. No TeV detection was reported after October 2008, until the source unexpectedly reappeared, once, at periastron \citep{2011ApJ...738....3A,2010ATel.2948....1O}. At the same time, 
a hard X-ray multi-year analysis \citep{2010MNRAS.408..642Z} and
a long-term X-ray campaign on \lsi\ using the \emph{Rossi X-ray Timing Explorer (RXTE)} has been conducted covering the whole extent of the LAT observations \citep[see][]{2010ApJ...719L.104T,2011ApJ...733...89L}. 
In addition, simultaneous and archival data from long-term monitoring of radio and H$\alpha$ emission is available for comparison in a multi-wavelength context.
In this work we present the results of the analysis of 30 months of LAT survey observations of both \lsi\ and
\ls. We investigate the long-term flux variations of the sources, as well as variations in the amplitude of
their orbital flux modulation, and we explore the possible spectral variability for both systems, finally putting and interpreting these observations in the context of the source behavior at other frequencies.


\section[data]{Observations and data reduction}\label{data}

The \fermi-LAT is an electron-positron pair production telescope, featuring solid state silicon trackers and cesium iodide
calorimeters, sensitive to photons from $\sim$20\,MeV to $>$300\,GeV \citep{2009ApJ...697.1071A}. It has a large $\sim$2.4~sr field of view (at 1\,GeV) and an effective area of $\sim$8000\,cm$^2$ for $>$1\,GeV.

\begin{figure*}[tbh]
    \begin{center}
    \includegraphics*[width=0.49\textwidth,angle=0,clip]{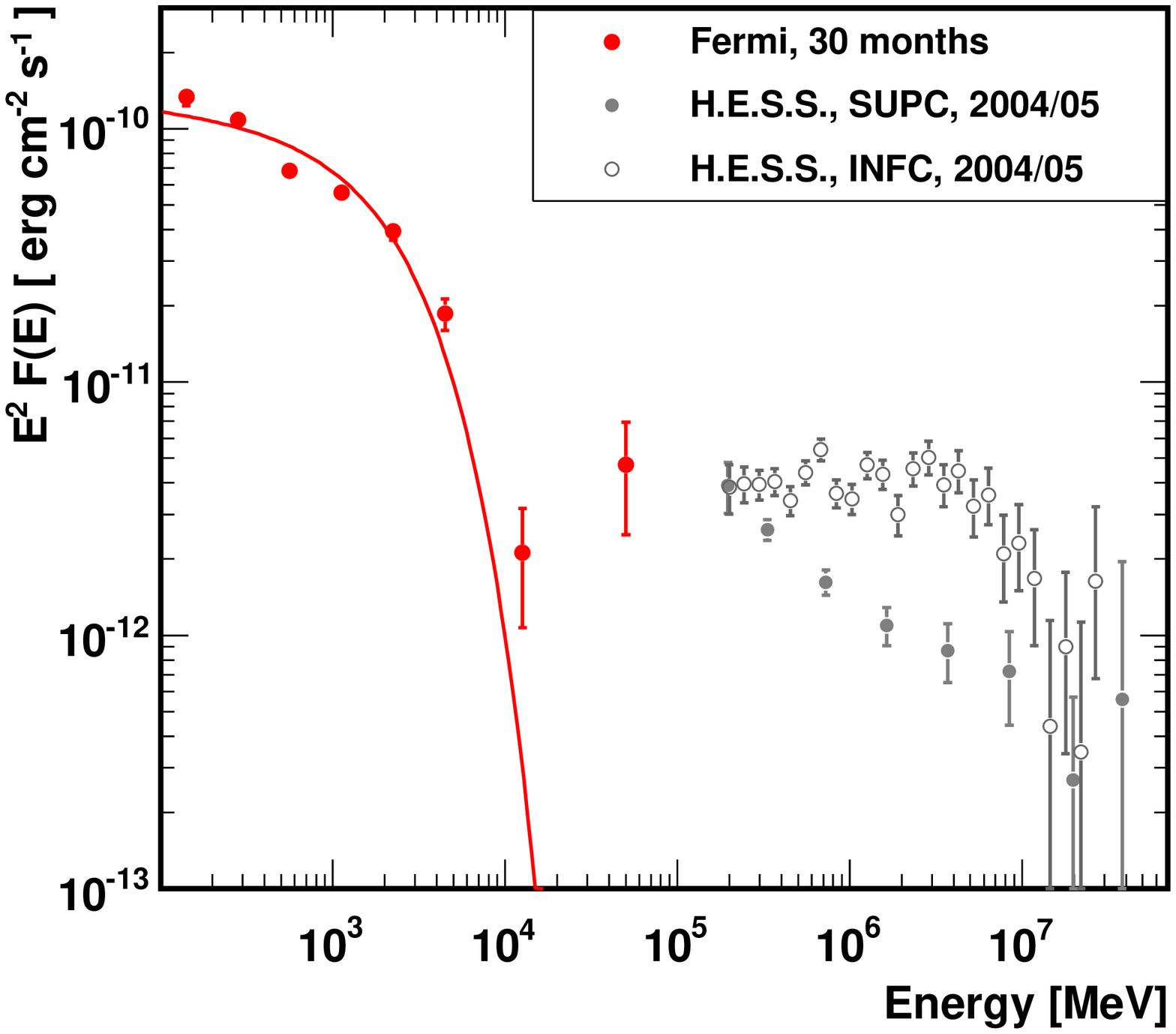}
    \includegraphics*[width=0.49\textwidth,angle=0,clip]{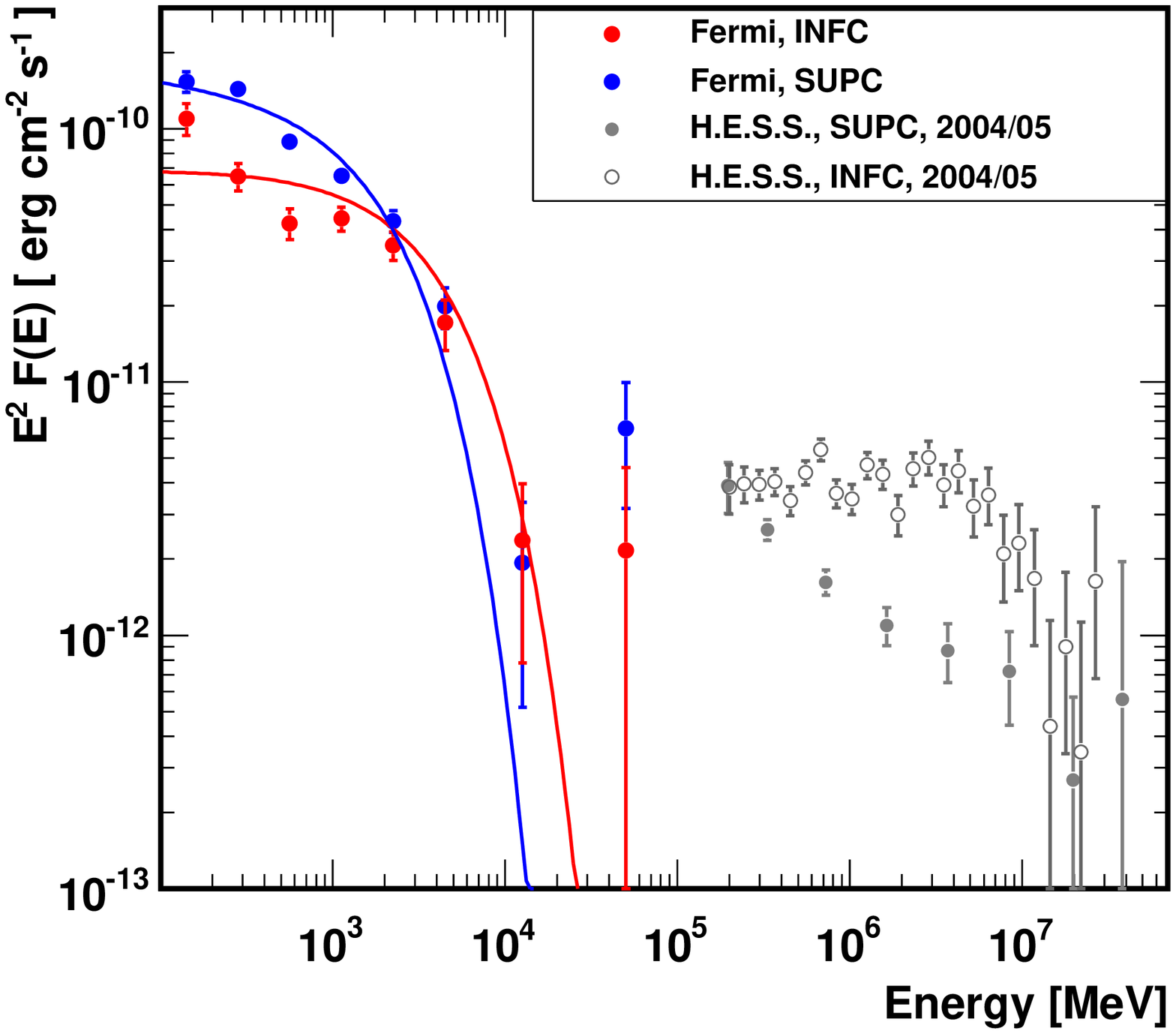}
\caption{\label{fig:LS5039_spectra}\textit{Left: }Spectral energy distribution of \ls\ as observed by the \fermi-LAT (red). The gray open and filled circles are H.E.S.S. spectra in INFC and SUPC, respectively \citep{2006A&A...460..743A} and are not simultaneous with the GeV measurements.
\textit{Right: }\fermi-LAT and H.E.S.S. spectra of \ls\ in SUPC (blue) and INFC (red).}
    \end{center}
\end{figure*}

\subsection{Dataset} 
The \fermi\ survey mode operations began on 2008 August 4; in this mode, the observatory is rocked north and south on alternate orbits to provide a more uniform coverage, so that every part of the sky is usually observed for $\sim$30 minutes every $\sim$3 hours. Therefore, the two sources of interest were monitored regularly without significant breaks, allowing us to draw a complete picture of their behavior in $\gamma$-rays over the last two years. The dataset used for this analysis spans 2008
August 4 through 2011 January 24.

The data were reduced and analyzed using the \textsc{Fermi Science Tools v9r20} package\footnote{See the Fermi Space Science Center (FSSC) website for details of the Science Tools: http://fermi.gsfc.nasa.gov/ssc/data/analysis/}.
The standard onboard filtering, event reconstruction, and classification were applied to the data \citep{2009ApJ...697.1071A}. The high-quality ``diffuse" event class was used together with
the ``Pass 6 v3 Diffuse" instrument response functions (IRFs). Time periods when the target source was observed at a zenith angle greater than 105$^\circ$ were excluded to limit contamination from Earth limb photons. Where required in the analysis, models for the Galactic diffuse emission (\textit{gll\_iem\_v02.fit}) and isotropic backgrounds (\textit{isotropic\_iem\_v02.txt}) were used\footnote{Descriptions of the models are available from the FSSC: http://fermi.gsfc.nasa.gov/ssc}.

\subsection{Spectral analysis methods}
The binned maximum-likelihood method of {\tt gtlike}, included in the ScienceTools, was used to determine the intensities and spectral parameters presented in this paper. We used all photons with energy $>$100\,MeV in a circular region of interest (ROI) of $10^{\circ}$ radius centered at the position of \lsi\ and \ls, respectively.
For source modeling, the 1FGL catalog \citep{2010ApJS..188..405A}, derived from 11 months of survey data, and the first \textit{Fermi} pulsar catalog \citep{2010ApJS..187..460A} were used; all sources within 15$^\circ$ of the ROI center were included.
The energy spectra of point sources included in the catalog within our ROI are modeled by a simple power law,
\begin{equation}
\frac{\mathrm{d}N}{\mathrm{d}E} = N_{0} \left(\frac{E}{E_0}\right)^{-\Gamma} 
\label{power_law}
\end{equation}
with the exception of known $\gamma$-ray pulsars, which were modeled by power laws with exponential cutoffs described by: 
\begin{equation}
\frac{\mathrm{d}N}{\mathrm{d}E} = N_{0} \left(\frac{E}{E_0}\right)^{-\Gamma} \mathrm{exp}\left[-\left(\frac{E}{E_{\mathrm{cutoff}}}\right)\right].
\label{exp_cut}
\end{equation}
The spectral parameters were fixed to the catalog values except for the sources within 3 degrees of the candidate location. For these latter sources, the flux normalization was left free. All of the spectral parameters of the two subject binaries were left free for the fit.  Source detection significance is determined using the Test Statistic value, $TS = -2\ \ln(L_{\rm 0}/L_{\rm 1})$ which compares the likelihood ratio of models including, e.g., an additional source, with the null-hypothesis of background only \citep{1996ApJ...461..396M}.

To estimate the systematic errors, which are mainly caused by uncertainties in the effective area and energy response of the LAT as well as background contamination, we use the so-called ``bracketing" IRFs. These are IRFs with effective areas that bracket those of our nominal IRF above and below by linearly connecting differences of (10\%, 5\%, 20\%) at log(E/MeV) of (2, 2.75, 4) respectively.

\subsection{Timing analysis methods}
Lightcurves are extracted using aperture photometry, taking an aperture radius
of 1$^\circ$ and using the {\tt gtbin} tool. The exposure correction is
performed with the tool {\tt gtexposure} assuming the spectral shape of the
source to be a power law with an exponential cutoff (see
\S\S~\ref{sec:LS_av_spec}, \ref{sec:lsi_av_spec}). These lightcurves are not
background subtracted.  The folded lightcurves shown in the subsequent sections
are derived by performing {\tt gtlike} fits for each phase bin. Therefore, all
of them are effectively background subtracted. We check that both methods for generating lightcurves,
aperture photometry and {\tt gtlike} fits, are
consistent with each other when the former lightcurves are
background-subtracted too.

The primary method of timing analysis employed searches for periodic modulation by calculating the weighted
periodogram of the lightcurve \citep{1976Ap&SS..39..447L, 1982ApJ...263..835S, 2007AIPC..921..548C}.  The lightcurve is constructed by summing, for each photon, the estimated
probability that the photon came from the source of interest.
The probability will be both spatially and spectrally dependent.
Because this technique allows for the correct weighting
of each photon it intrinsically improves the signal-to-noise and allows the use of a larger aperture.
This method has successfully been applied to increasing the
LAT sensitivity for the detection of pulsars \citep{2011ApJ...732...38K}. However,
in the basic form of this technique, the weight for any particular
energy/position is fixed. This means that changes in source
brightness will not be reflected in the weights and can result
in incorrect probabilities. The calculation of probabilities was performed using the tool {\tt gtsrcprob}
and the same source model file derived from the 1FGL catalog and used in the spectral analysis.  Since the exposure of the time bins was variable, the contribution of each time bin to the power spectrum was weighted based on its relative exposure. Period errors are calculated using the method of \citet{1986ApJ...302..757H}.


\section{\ls\ Results}\label{LS5039}

\ls\ is located in a complicated region toward the inner Galaxy with high Galactic diffuse emission and many surrounding $\gamma$-ray sources. In particular, the LAT detected a bright ($8.7 \times 10^{-7}~{\rm ph}~{\rm cm}^{-2}~{\rm s}^{-1}$ above 100~MeV) $\gamma$-ray pulsar, PSR~J1826$-$1256, $\sim2^\circ$ away from \ls. Following the analysis performed in the earlier LAT paper \citep{2009ApJ...706L..56A} we discarded events whose arrival times correspond to the peaks 
of the pulsar cycle of PSR~J1826$-$1256 in order to minimize the contamination from the pulsar. 
The excluded pulse phase of PSR~J1826$-$1256 is $0.05 < \phi_p < 0.2$ and $0.625 < \phi_p < 0.775$ (see Fig. 37 in \citet{2011ApJS..194...17R}), which 
results in a loss of 30\% exposure on LS 5039. To account for the loss, a scaling factor of $1/0.7$ is multiplied to fluxes obtained with maximum likelihood fits.

\subsection{Orbitally averaged spectrum}\label{sec:LS_av_spec}

The orbitally-averaged spectrum of \ls\ was initially investigated by fitting a
power law and a power law with an exponential cutoff to the data.  We compare
two models utilizing the likelihood ratio test \citep{1996ApJ...461..396M},
i.e. for the ratio $2\times\Delta log(Likelihood)$ we assume a
$\chi^{2}$-distribution to calculate the probabilities taking into account the
corresponding degrees of freedom \citep{Eadie1998}.
In this case, the significance of 
a spectral cutoff was assessed by comparing the likelihood 
ratio between the power law and cutoff power law cases, which is $-2\ \ln(L_{\rm PL}/L_{\rm cutoff}) = 94.9$, where 
$L_{\rm PL}$ and $L_{\rm cutoff}$ are the likelihood value obtained for the spectral fits with a power law and a cutoff power law, respectively.  This indicates that the simple power law model is rejected at the 9.7$\sigma$ level in favor of the cutoff power law.  The best-fit parameters for the cutoff power law model are 
$\Gamma = 2.06 \pm 0.06_{\mathrm{stat}} \pm 0.11_{\mathrm{syst}}$
and 
$E_{\rm cutoff} = 2.2 \pm 0.3_{\mathrm{stat}} \pm 0.5_{\mathrm{syst}}$~GeV 
with a flux of 
$F_{> 100~{\rm MeV}} = (6.1 \pm 0.3_{\mathrm{stat}} \pm 2.1_{\mathrm{syst}}) \times 10^{-7}~{\rm cm}^{-2}~{\rm s}^{-1}$ 
integrated above 100~MeV. Using a cutoff power-law spectral model, the maximum likelihood fit yields a test statistic of TS = 1623 for the \ls\ detection; equivalent to $\sim40~\sigma$.
We also tried a broken power-law spectral model for \ls\ in addition to an exponentially cutoff power law. We found the broken power law gives lower TS values than the exponentially cutoff power-law case.

Spectral points in each energy band were obtained by dividing the dataset into separate energy bins and 
performing maximum likelihood fits for each of them. The resulting spectral energy distribution (SED) is plotted 
in Figure~\ref{fig:LS5039_spectra} together with the best-fit cutoff power law model. Interestingly, the SED shows significantly higher flux (one spectral data point) at $\gtrsim 10$~GeV than the expected flux from the best-fit cutoff power law, possibly 
suggesting another component at high energies.

One idea explored for these sources (especially, for \ls,  see e.g., Torres 2010) is that 
the $\gamma$-ray emission could be understood as
having two components: one would be the magnetospheric GeV emission
from a putative pulsar and the other from the inter-wind region or
from the pulsar wind zone. The latter would be unpulsed and would vary with the
orbital phase, the former would be steady and pulsed. The current data for \ls\ would indeed allow for
this possibility, especially because of the possible high energy component found in the GeV spectrum. 

To test the significance of the additional component, we added to the model a
power-law source at the location of \ls\ in addition to the cutoff power law
source.
Model A is a power law with a cutoff and model B a power law with a
cutoff plus an additional power law. Thus, model B has two more free parameters
compared to model A. According to the likelihood ratio test, the
probability of incorrectly rejecting model A is $6.1\times10^{-6}$ ($4.5\sigma$). 

The best-fit parameters for the putative additional component are 
$\Gamma = 1.6 \pm 0.4_{\mathrm{stat}} \pm 0.3_{\mathrm{syst}}$ and $F_{> 10~{\rm GeV}} = (1.6 \pm 0.6_{\mathrm{stat}} \pm 0.7_{\mathrm{syst}}) \times 10^{-10}~{\rm cm}^{-2}~{\rm s}^{-1}$.  
The addition of the high-energy component slightly affects the parameters of the cutoff power law. The best-fit parameters for the latter are $\Gamma = 2.02 \pm 0.06_{\mathrm{stat}} \pm 0.10_{\mathrm{syst}}$ and 
$E_{\rm cutoff} = 2.0 \pm 0.3_{\mathrm{stat}} \pm 0.4_{\mathrm{syst}}$~GeV with a flux of 
$F_{> 100~{\rm MeV}} = (6.0 \pm 0.3_{\mathrm{stat}} \pm 1.9_{\mathrm{syst}} \times 10^{-7}~{\rm cm}^{-2}~{\rm s}^{-1}$ integrated above 100~MeV.
  
\begin{figure}[tbh]
    \begin{center}
    \includegraphics*[width=0.49\textwidth,angle=0,clip]{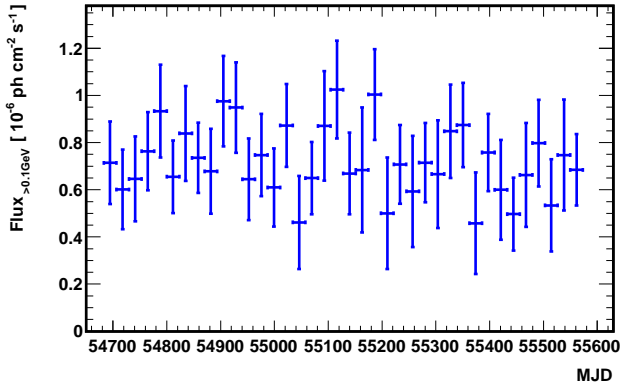}
\caption{\label{fig:whole_data_lc_ls5039}
Light curve of \ls\ in time bins of six orbital cycles between 100\,MeV-300\,GeV. One orbit is 3.90532 days long.}
    \end{center}
\end{figure}

\begin{figure}
        \begin{center}
    \includegraphics*[width=0.49\textwidth,angle=0,clip]{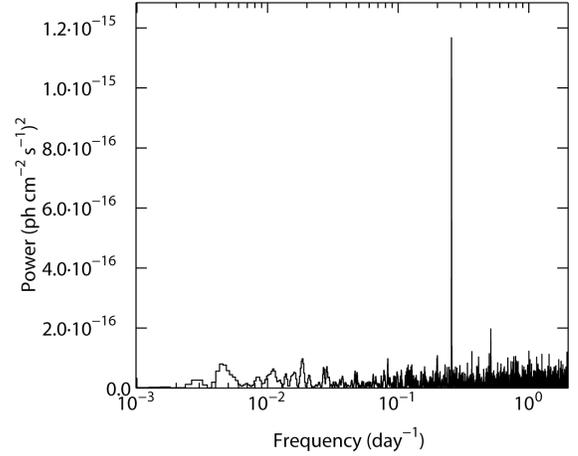}
\caption{\label{fig:whole_data_lc_ls5039_timing} Lomb-Scargle power spectrum of the whole GeV data set on \ls.} 
      \end{center}
\end{figure}

\subsection{Phase-resolved analysis}

Following the H.E.S.S. analysis by \citet{2006A&A...460..743A} as well as  the previous one, the whole dataset was divided into two orbital intervals: superior conjunction (SUPC; $\phi < 0.45$ and $0.9 < \phi$) and inferior conjunction (INFC; $0.45 < \phi < 0.9$).  The SUPC and INFC data were analyzed in the same way as the orbitally averaged data. 
Being consistent with our previous paper, the power-law assumption for the SUPC spectrum is 
rejected with $-2\ \ln(L_{\rm PL}/L_{\rm cutoff}) = 81.2$, or at a rejection significance 
of $\sim 9\sigma$.  The best-fit parameters are $\Gamma = 2.07 \pm 0.07_{\mathrm{stat}} \pm 0.08_{\mathrm{syst}}$, 
$E_{\rm cutoff} = 1.9 \pm 0.3_{\mathrm{stat}} \pm 0.3_{\mathrm{syst}}$~GeV, and $F_{> 100~{\rm MeV}} = (7.8 \pm 0.4_{\mathrm{stat}} \pm 1.9_{\mathrm{syst}}) \times 10^{-7}~{\rm cm}^{-2}~{s}^{-1}$.
 
Although a single power law was not rejected for INFC in our previous analysis using 10 months of data \citep{2009ApJ...706L..56A}, a cutoff power law is preferred also for INFC with the present dataset.  The likelihood ratio for the INFC data is $-2\ \ln(L_{\rm PL}/L_{\rm cutoff}) = 21.7$, which corresponds to 4.7$\sigma$. The parameters for the INFC spectrum are $\Gamma = 1.99 \pm 0.13_{\mathrm{stat}} \pm 0.07_{\mathrm{syst}}$, $E_{\rm cutoff} = 2.6 \pm 0.7_{\mathrm{stat}} \pm 0.9_{\mathrm{syst}}$~GeV, and $F_{> 100~{\rm MeV}} = (3.9 \pm 0.4_{\mathrm{stat}} \pm 1.5_{\mathrm{syst}}) \times 10^{-7}~{\rm cm}^{-2}~{\rm s}^{-1}$.
Therefore, the SUPC and INFC spectral shapes are completely consistent with one another within the errors. The only difference is the normalization and hence the total flux.
On the other hand, the spectrum for INFC (red points in the right panel of
Figure~\ref{fig:LS5039_spectra}) seems to exhibit additional structure below 1\,GeV.
The limited statistics and the large contribution of diffuse
emission at low energies, however,
prevents solid conclusions on whether a more complicated fit
(e.g.\ a double broken power law or a broken power law with a cutoff) would be preferred.
 
We also searched for emission from the high-energy component in the SUPC and INFC spectra. 
However, the $TS$ of the additional components compared with a power law with exponential cutoff are only 13.6 and 10.9 for SUPC and INFC, respectively, and do not confirm a second spectral component.
The SUPC and INFC SEDs were obtained using the same method as the orbitally averaged spectra and are plotted in the right panel of Figure~\ref{fig:LS5039_spectra}. 
 
\subsection{Lightcurve}

Figure~\ref{fig:whole_data_lc_ls5039} shows the lightcurve for  \ls\ over 30 months derived by performing {\tt gtlike} fits on time bins which contain 6 orbital cycles each.  The lightcurve for \ls\ does not show any significant flux changes. Constructing the periodogram of the weighted photon lightcurve yields a significant detection of a periodicity at 3.90532 $\pm$ 0.0008 days. This is consistent with the known orbital period of \ls. The Lomb-Scargle power spectrum of \ls\ is shown in Figure \ref{fig:whole_data_lc_ls5039_timing}.  The stability of the orbital modulation was investigated and no significant variation in the modulation fraction as a function of time was found.


\section{LS I +61$^\circ$303 Results}\label{LSI}

\subsection{Orbitally-averaged spectral analysis}\label{sec:lsi_av_spec}

\begin{figure}[tbh]
    \begin{center}
    \includegraphics*[width=0.49\textwidth,angle=0,clip]{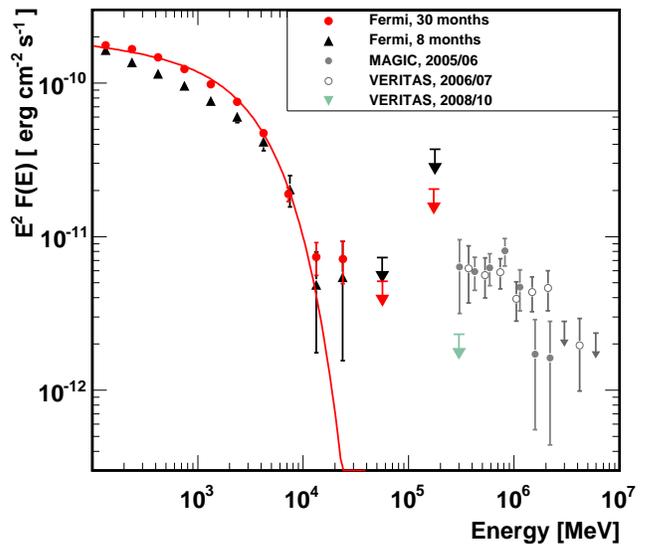}
\caption{\label{fig:whole_data}The overall 30-months \lsi\ spectrum (red) in comparison with the earlier published one (black) over 8 months is shown. 
TeV data points taken by MAGIC and VERITAS are shown in gray. They are not simultaneously taken with the GeV data. The VERITAS  upper limit (in green) is.
} 
    \end{center}
\end{figure}

We have derived the spectrum of orbitally-averaged LAT data, i.e., without any selection criteria (cuts) concerning the orbital phase, for the \lsi\ system.
The spectral points and corresponding best-fit using the updated dataset described in \S~\ref{data} are shown in Figure
\ref{fig:whole_data}, together with previously derived results from the LAT and
TeV observations. Two sets of TeV data are plotted: we show the
non-simultaneous data points obtained by the Cherenkov
telescope experiments MAGIC and VERITAS. (These data correspond to phases around 0.6--0.7 and represent several orbits observed in the period 2006--2008, before \fermi\ was launched). 
Additionally, we show the latest measurements performed by VERITAS, which
established a 99\% C.L. upper limit.\footnote{We derive this differential upper limit by using the VERITAS-reported integral flux upper limit for phases 0.6--0.7 \citep{2011ApJ...738....3A} assuming a differential spectral slope of 2.6.} The new VERITAS upper limit
spans several orbits during which, simultaneously with our LAT data, no detection was achieved. The
LAT data along the whole orbit are still best described by a power law with an
exponential cutoff. 
The $TS$~value for a source emitting $\gamma$-rays at the
position of \lsi\ with an SED described by a power law with an exponential
cutoff is highly significant.
The relative $TS$~value comparing a fit with a power law and a fit with a
power law plus an exponential cutoff clearly favors the latter, at the
$\sim$20$\sigma$ level.
The photon index found is $\Gamma = 2.07 \pm 0.02_{\mathrm{stat}} \pm 0.09_{\mathrm{syst}}$; the flux above
100\, MeV is $(0.95 \pm 0.01_{\mathrm{stat}} \pm 0.07_{\mathrm{syst}}) \times 10^{-6} \, \mathrm{ph\, cm^{-2}\, s^{-1}}$, and the cutoff energy is $3.9 \pm 0.2_{\mathrm{stat}} \pm 0.7_{\mathrm{syst}}\, \mathrm{GeV}$.  
Results for the obtained $TS$ values for each fit to different datasets are
listed in Table~\ref{table:TS} and all fit parameters obtained for the exponentially cutoff power law models are listed in
Table~\ref{table:fit}.

\begin{figure}[tbh]
    \begin{center}
    \includegraphics*[width=0.49\textwidth,angle=0,clip]{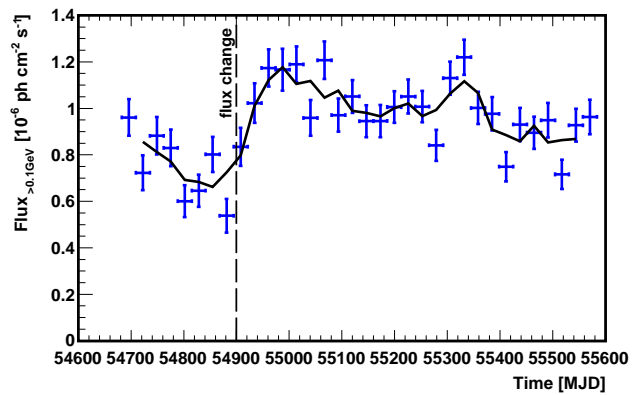}
\caption{\label{fig:orbital_lc_time_LSI}Lightcurve of LS I +61$^\circ$303. Each point represents one orbit of 26.496 days, whereas the black solid line represents the 3-bin smoothed light curve. The black dashed line marks the moment of the flux change in March 2009. 
} 
    \end{center}
\end{figure}

\begin{figure}[tbh]
    \begin{center}
   \includegraphics*[width=0.49\textwidth,angle=0,clip]{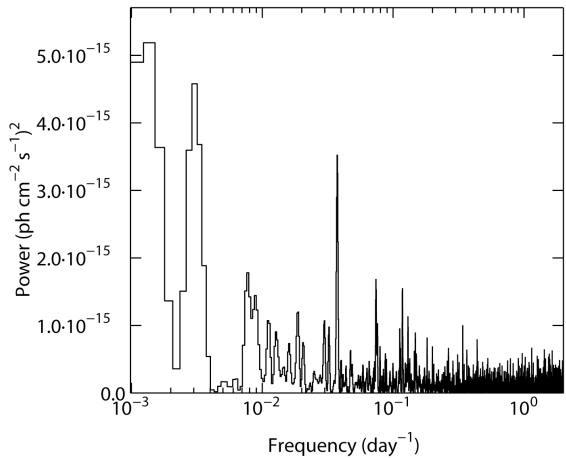}
\caption{\label{fig:orbital_lc_time_LSI_timing}
The Lomb-Scargle periodogram of the whole \lsi\ GeV data set. The orbital period is clearly visible. } 
    \end{center}
\end{figure}

\begin{figure*}[tbh]
    \begin{center}
    \includegraphics*[width=0.49\linewidth,angle=0,clip]{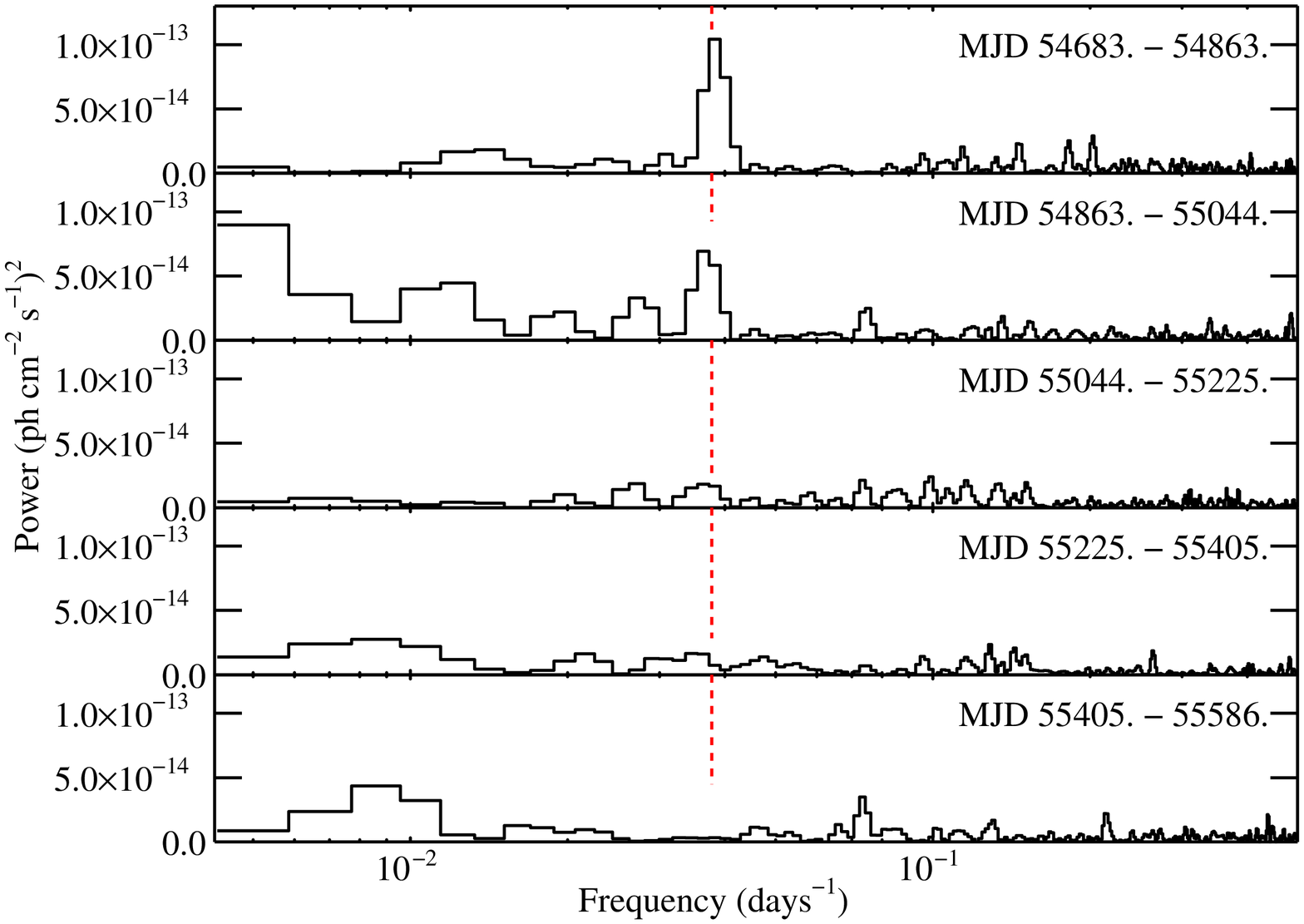}
    \includegraphics*[width=0.49\linewidth,angle=0,clip]{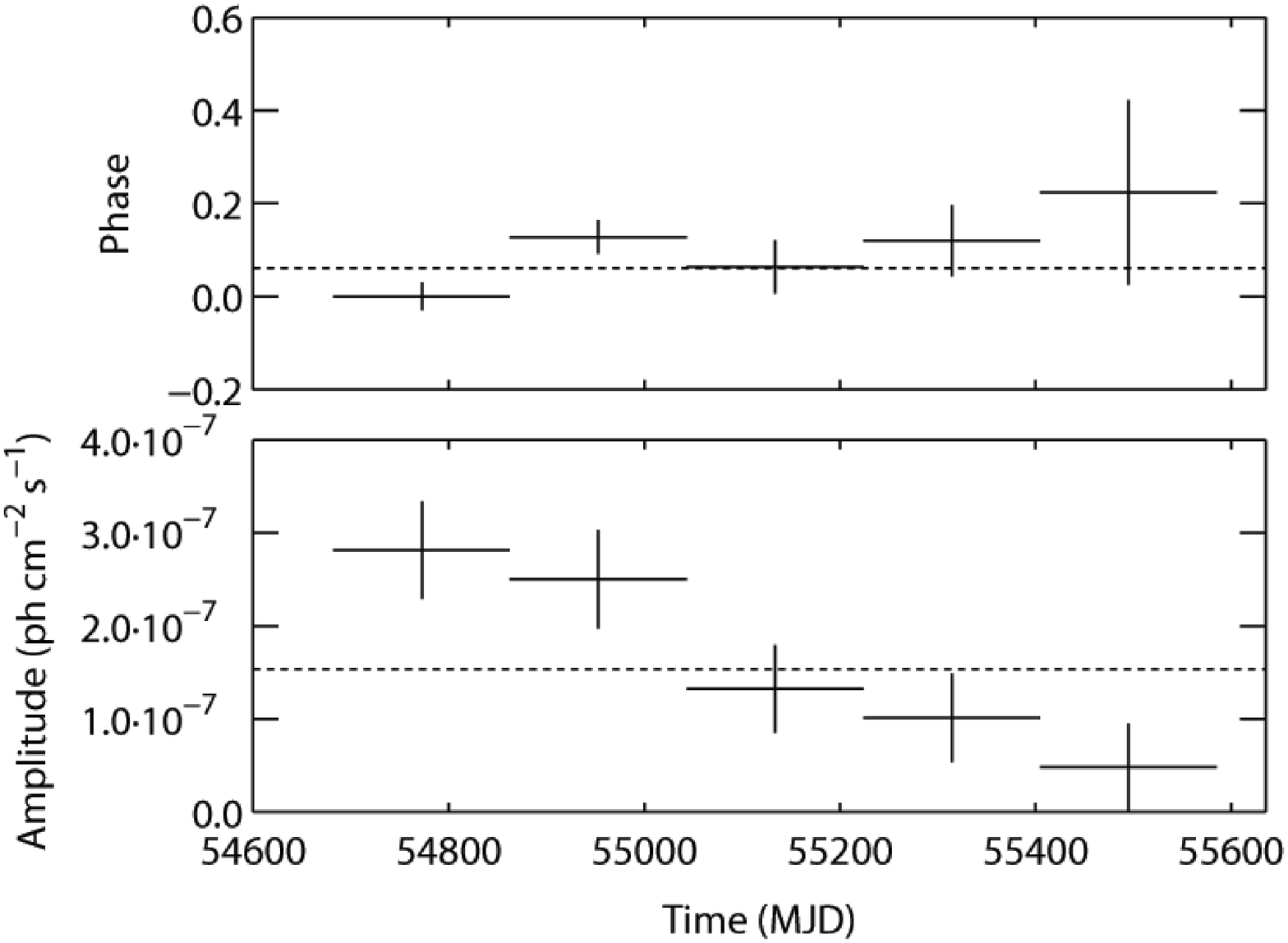}
\caption{\label{fig:lsi_varying_power} \textit{Left:} The Lomb-Scargle periodograms of the \lsi\ 30 months lightcurve  split into five consecutive segments; the earliest is at the top. The red dashed line indicates the known orbital period.
\textit{Right:} The change in the amplitude and phase of the orbital flux modulation calculated by fitting a sine wave to each of the lightcurve  segments.
} 
    \end{center}
\end{figure*}

\begin{table*}[tbh]
\scriptsize
\begin{center}
\caption{\label{table:TS}$TS$ values for \lsi\ for different spectral shapes (see Section~\ref{phase} for details)}
\begin{tabular}{ l c c c }
\hline
Data set                             & Power law + exponential cutoff    &  Power law & Broken power law\\
                                     & TS                                & TS         & TS               \\ \hline \hline
30 months of data                    & 23995                           & 23475    & 23970                       \\
Data before March 2009               & 3404                            & 3314     & 3415                         \\
Data after March 2009                & 20714                           & 20283    & 20699                        \\
Inferior conjunction (geometrically) & 12548                           & 12326    & 12512                          \\
Superior conjunction (geometrically) & 11711                           & 11422    & 11700                           \\
Inferior conjunction (angle cut)     & 6670                            & 6562     & 6665                           \\
Superior conjunction (angle cut)     & 6083                            & 5986     & 6063                            \\
Periastron                           & 11656                           & 11450    & 11636                         \\
Apastron                             & 12377                           & 12059    & 12361                              \\
\hline
\end{tabular}
\end{center}
\end{table*}

\begin{table*}[bth]
\scriptsize
\begin{center}
\caption{\label{table:fit}Parameters for \lsi\ from spectral fitting with power law with exponential cutoff (see Section~\ref{phase} for details)}
\begin{tabular}{ l c c c}
\hline
Data set                             & Photon index $\Gamma$                                    & Cutoff energy                                       & Flux $>$100\,MeV                                              \\
                                     &                                                          & [GeV]                                               & $[\times 10^{-6}$ ph\, cm$^{-2}$\, s$^{-1}]$       \\ \hline \hline
First 8 months of data               & $2.21 \pm 0.04_{(\mathrm{stat}} \pm 0.06_{\mathrm{syst}}$ & $6.3 \pm 1.1_{\mathrm{stat}} \pm 0.4_{\mathrm{syst}}$ & $0.82 \pm 0.03_{\mathrm{stat}} \pm 0.07_{\mathrm{syst}}$  \\
30 months of data                    & $2.07 \pm 0.02_{\mathrm{stat}} \pm 0.09_{\mathrm{syst}}$ & $3.9 \pm 0.2_{\mathrm{stat}} \pm 0.7_{\mathrm{syst}}$ & $0.95 \pm 0.01_{\mathrm{stat}} \pm 0.07_{\mathrm{syst}}$   \\
Data before March 2009               & $2.08 \pm 0.04_{\mathrm{stat}} \pm 0.09_{\mathrm{syst}}$ & $4.0 \pm 0.6_{\mathrm{stat}} \pm 0.7_{\mathrm{syst}}$ & $0.75 \pm 0.03_{\mathrm{stat}} \pm 0.07_{\mathrm{syst}}$   \\
Data after March 2009                & $2.07 \pm 0.02_{\mathrm{stat}} \pm 0.09_{\mathrm{syst}}$ & $3.9 \pm 0.3_{\mathrm{stat}} \pm 0.7_{\mathrm{syst}}$ & $1.00 \pm 0.01_{\mathrm{stat}} \pm 0.07_{\mathrm{syst}}$   \\
Inferior conjunction (geometrically) & $2.14 \pm 0.02_{\mathrm{stat}} \pm 0.09_{\mathrm{syst}}$ & $4.0 \pm 0.4_{\mathrm{stat}} \pm 0.7_{\mathrm{syst}}$ & $1.07 \pm 0.02_{\mathrm{stat}} \pm 0.07_{\mathrm{syst}}$   \\
Superior conjunction (geometrically) & $2.02 \pm 0.03_{\mathrm{stat}} \pm 0.09_{\mathrm{syst}}$ & $3.9 \pm 0.3_{\mathrm{stat}} \pm 0.7_{\mathrm{syst}}$ & $0.85 \pm 0.02_{\mathrm{stat}} \pm 0.07_{\mathrm{syst}}$   \\
Inferior conjunction (angle cut)     & $2.17 \pm 0.03_{\mathrm{stat}} \pm 0.09_{\mathrm{syst}}$ & $4.1 \pm 0.5_{\mathrm{stat}} \pm 0.7_{\mathrm{syst}}$ & $1.11 \pm 0.03_{\mathrm{stat}} \pm 0.07_{\mathrm{syst}}$   \\
Superior conjunction (angle cut)     & $2.15 \pm 0.03_{\mathrm{stat}} \pm 0.09_{\mathrm{syst}}$ & $5.0 \pm 0.7_{\mathrm{stat}} \pm 0.7_{\mathrm{syst}}$ & $0.91 \pm 0.02_{\mathrm{stat}} \pm 0.07_{\mathrm{syst}}$   \\
Periastron                           & $2.14 \pm 0.02_{\mathrm{stat}} \pm 0.09_{\mathrm{syst}}$ & $4.1 \pm 0.4_{\mathrm{stat}} \pm 0.7_{\mathrm{syst}}$ & $1.01 \pm 0.02 _{\mathrm{stat}} \pm 0.07_{\mathrm{syst}}$  \\
Apastron                             & $2.01 \pm 0.03_{\mathrm{stat}} \pm 0.09_{\mathrm{syst}}$ & $3.7 \pm 0.3_{\mathrm{stat}} \pm 0.7_{\mathrm{syst}}$ & $0.90 \pm 0.02_{\mathrm{stat}} \pm 0.07_{\mathrm{syst}}$   \\
\hline
\end{tabular}
\end{center}
\end{table*}

Figure~\ref{fig:whole_data} shows that the data point at 30\,GeV deviates from the model by more than $3\sigma$ (power law with cutoff, red line). Although in our representation it is only one point, it is in itself significant, with a $TS$ value of 67 corresponding to $\sim$8$\sigma$. 
Therefore, and similarly to the case of \ls, with the caveat of having only one point determined in the SED beyond the results of the fitted spectral model,
we investigate the possible presence of a second component at high energies.
As in the case for \ls,  we use the likelihood ratio test to compare two models: Model A is a power law with a cutoff and model B a power law with a cutoff plus an additional power law. According to this test, the probability of incorrectly rejecting model A is $5.7\times10^{-15}$ ($7.8\sigma$).
The $TS$ value for this extra power law component as a whole is 172, larger than in the case of \ls, and its parameters are  $\Gamma = 2.5 \pm 0.3_{\mathrm{stat}}$ and $F_{> 10~{\rm GeV}} = (3.5 \pm 0.6_{\mathrm{stat}}) \times 10^{-10}~{\rm cm}^{-2}~{\rm s}^{-1}$.
The addition of the high-energy component affects the parameters of the cutoff power law. The best-fit parameters for the latter, when including the former, are $\Gamma = 2.00 \pm 0.03_{\mathrm{stat}}$ and 
$E_{\rm cutoff} = 2.7 \pm 0.3_{\mathrm{stat}}$~GeV with a flux of $F_{> 100~{\rm MeV}} = (0.88 \pm 0.08_{\mathrm{stat}}) \times 10^{-6}~{\rm cm}^{-2}~{\rm s}^{-1}$. 
Compared with the cutoff energy of $3.9 \pm 0.2$~GeV we obtain after fitting only a power law with an exponential cutoff to the data, the cutoff energy decreases when the additional high-energy power-law component is introduced.

An alternative model to accommodate the deviating high-energy point in the spectrum is a fit 
with a broken power-law. This is further discussed in Section~\ref{different_fits}.

\begin{figure*}[tbh]
    \begin{center}
   \includegraphics*[width=0.32\textwidth,angle=0,clip]{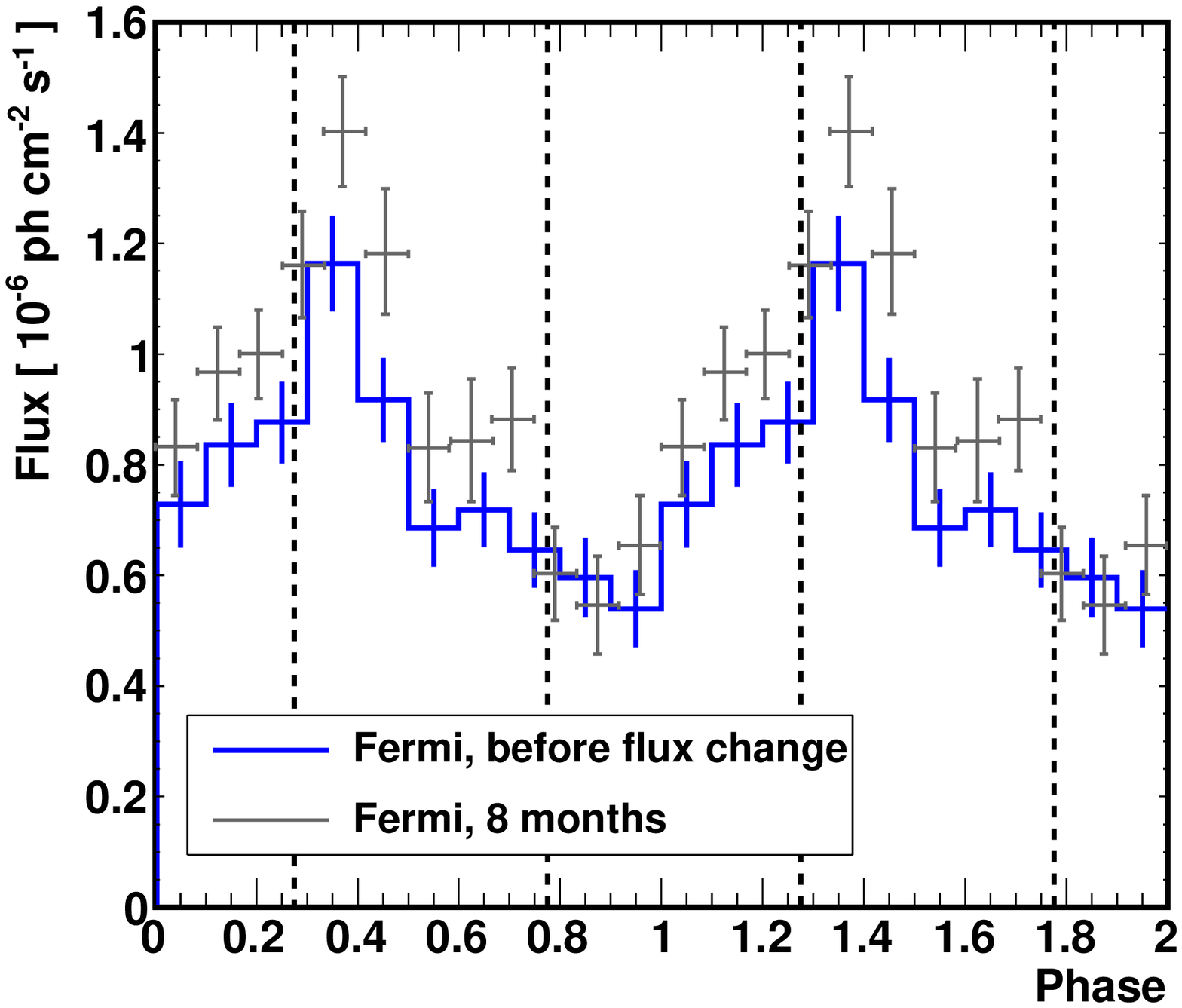}
    \includegraphics*[width=0.32\textwidth,angle=0,clip]{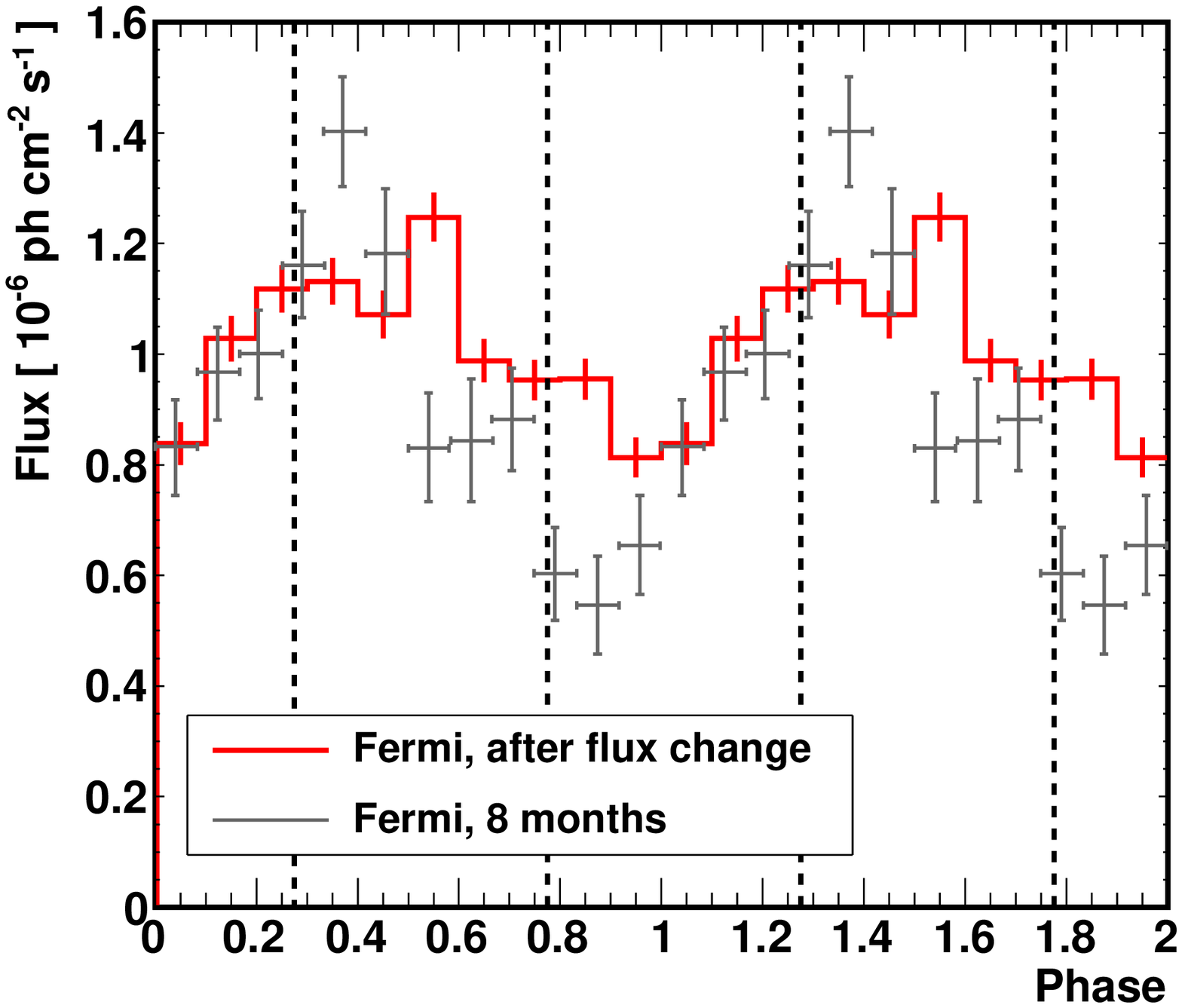}
\includegraphics*[width=0.32\textwidth,angle=0,clip]{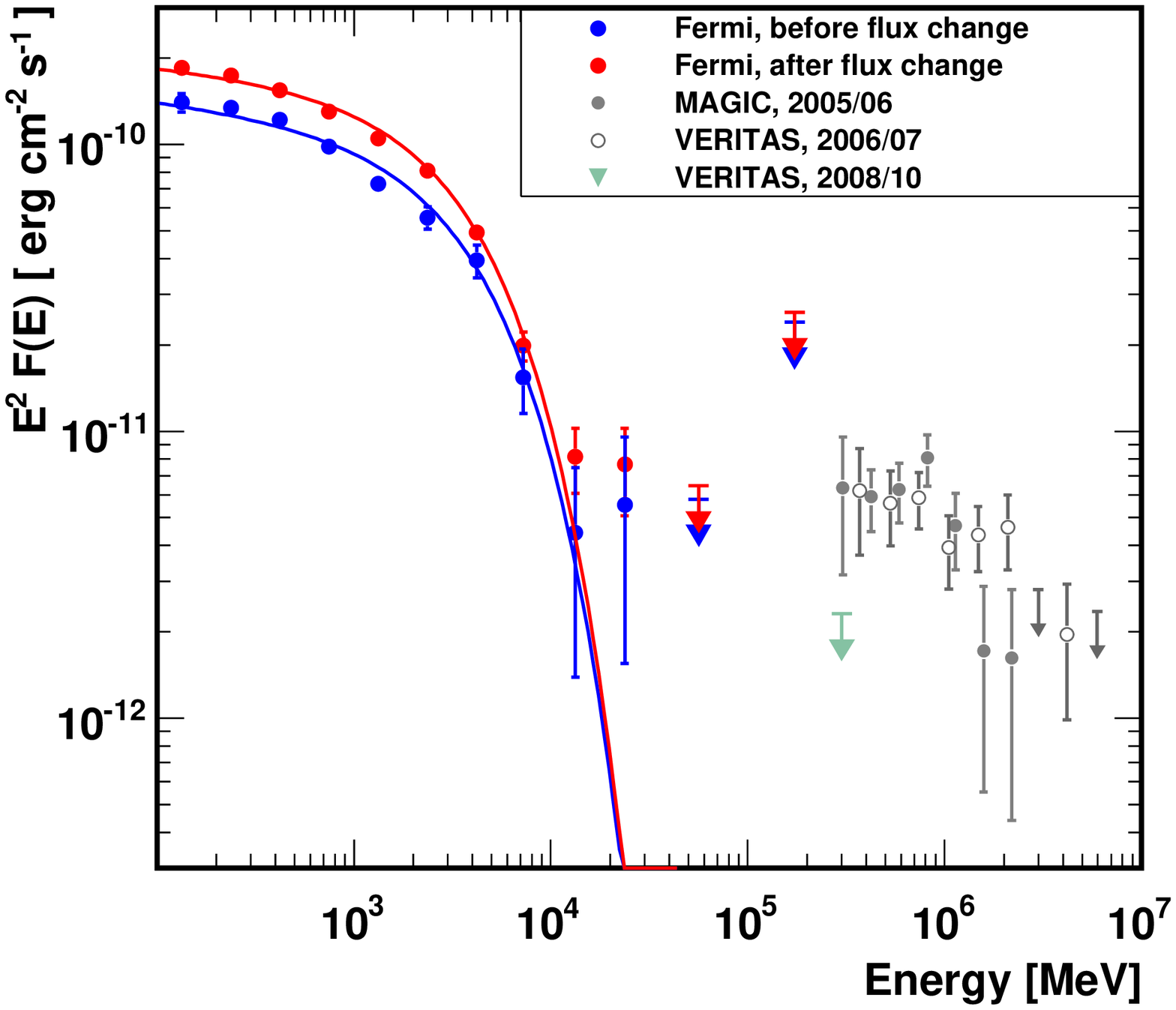}
\caption{\label{fig:before_after} \textit{Left: }Folded lightcurve (100\,MeV-300\,GeV) of \lsi\  before the flux change (blue), when the modulation is still clearly visible, compared with the earlier published 8-months dataset shown in gray. \textit{Middle: } Folded lightcurve (100\,MeV-300\,GeV) after the flux change, in March 2009 (red). The modulation gets fainter. For comparison, the previous published lightcurve is also plotted in gray.  \textit{Right: } Comparison of the spectra derived before (blue) and after (red) the flux change in March 2009. }
    \end{center}
\end{figure*} 
\begin{figure*}[ht]
    \begin{center}
    \includegraphics*[width=0.32\textwidth,angle=0,clip]{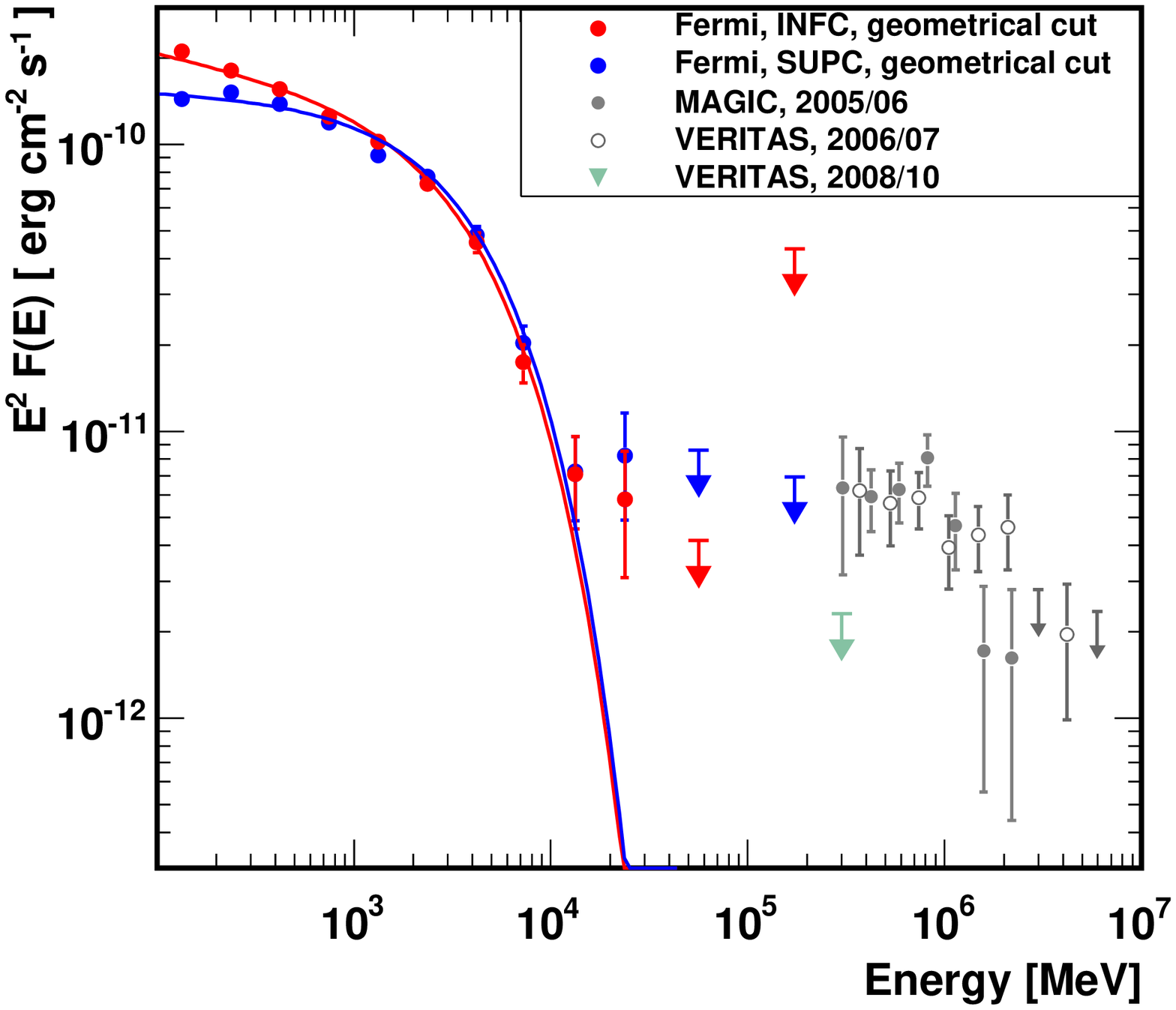}
    \includegraphics*[width=0.32\textwidth,angle=0,clip]{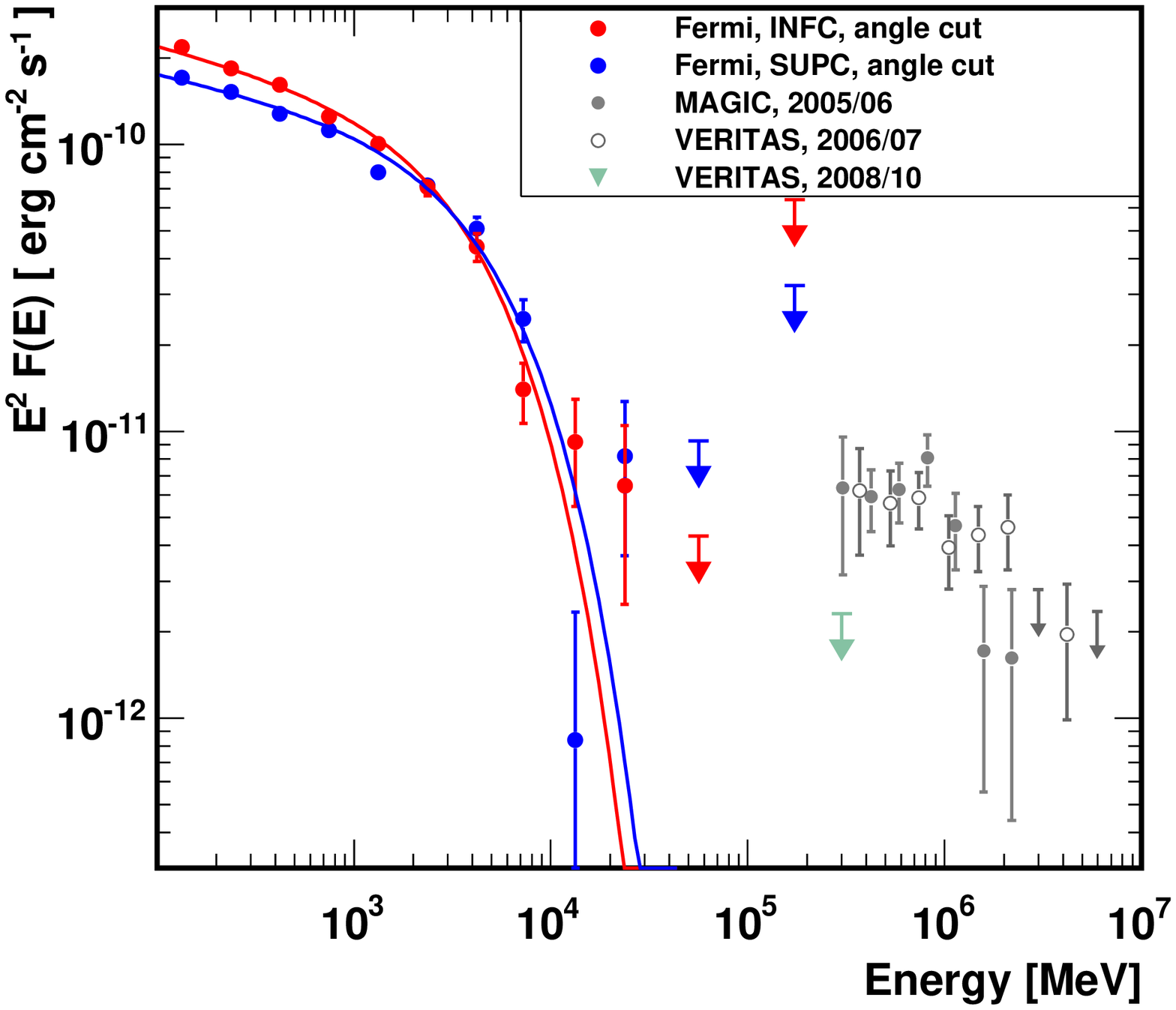}
    \includegraphics*[width=0.32\textwidth,angle=0,clip]{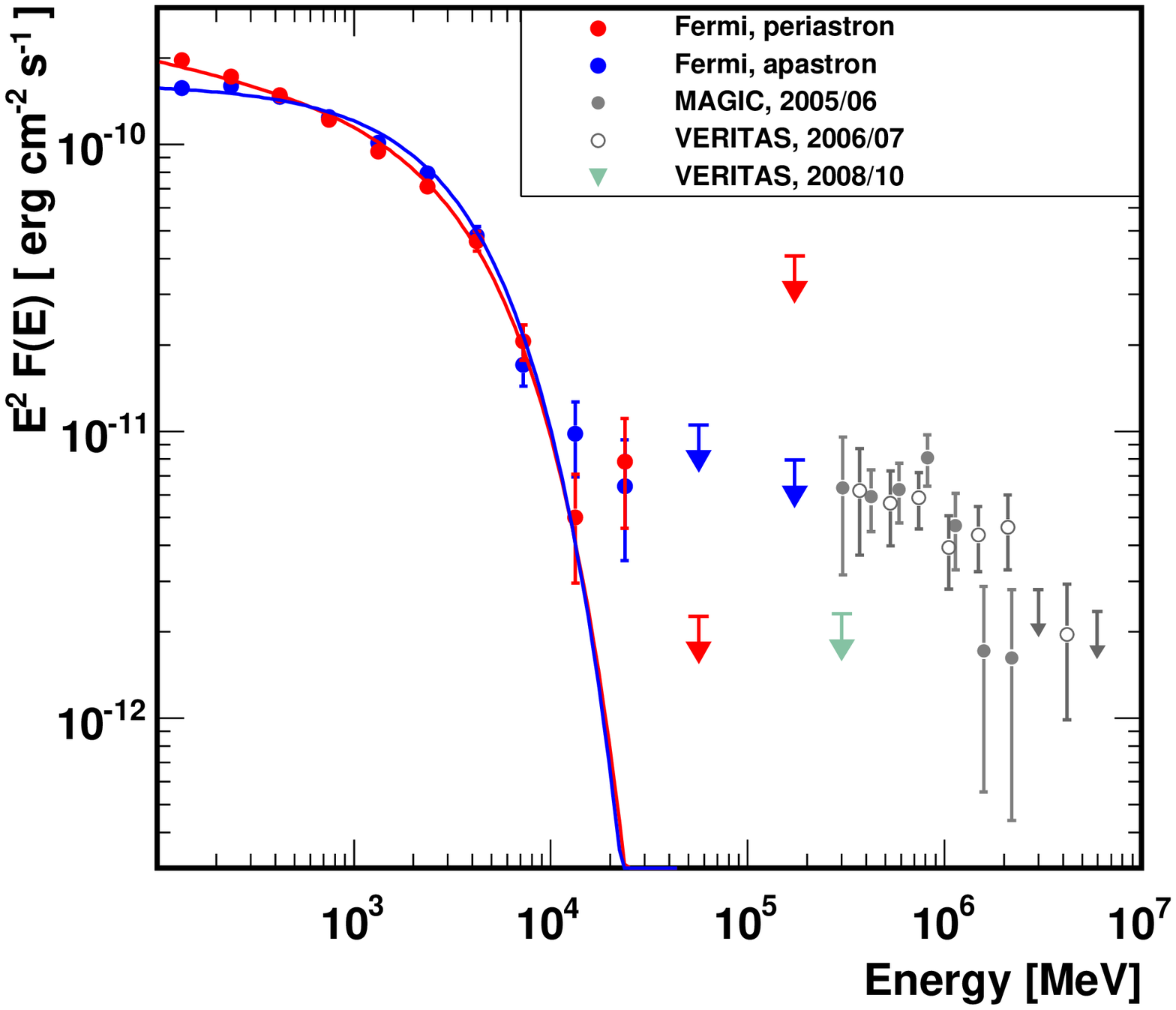}
\caption{\label{fig:orbital_cuts}Spectra of \lsi\ for different phase bins during the orbit are shown. \textit{Left: }Spectra for the geometrical cut of the orbit in superior (blue) and inferior (red) conjunction. \textit{Middle: }Spectra for the angle based cut in superior and inferior conjunction. \textit{Right: }Spectra derived for periastron (red) and apastron (blue).} 
    \end{center}
\end{figure*}

\begin{figure}[tbh]
    \begin{center}
    \includegraphics*[width=0.98\linewidth,angle=0,clip]{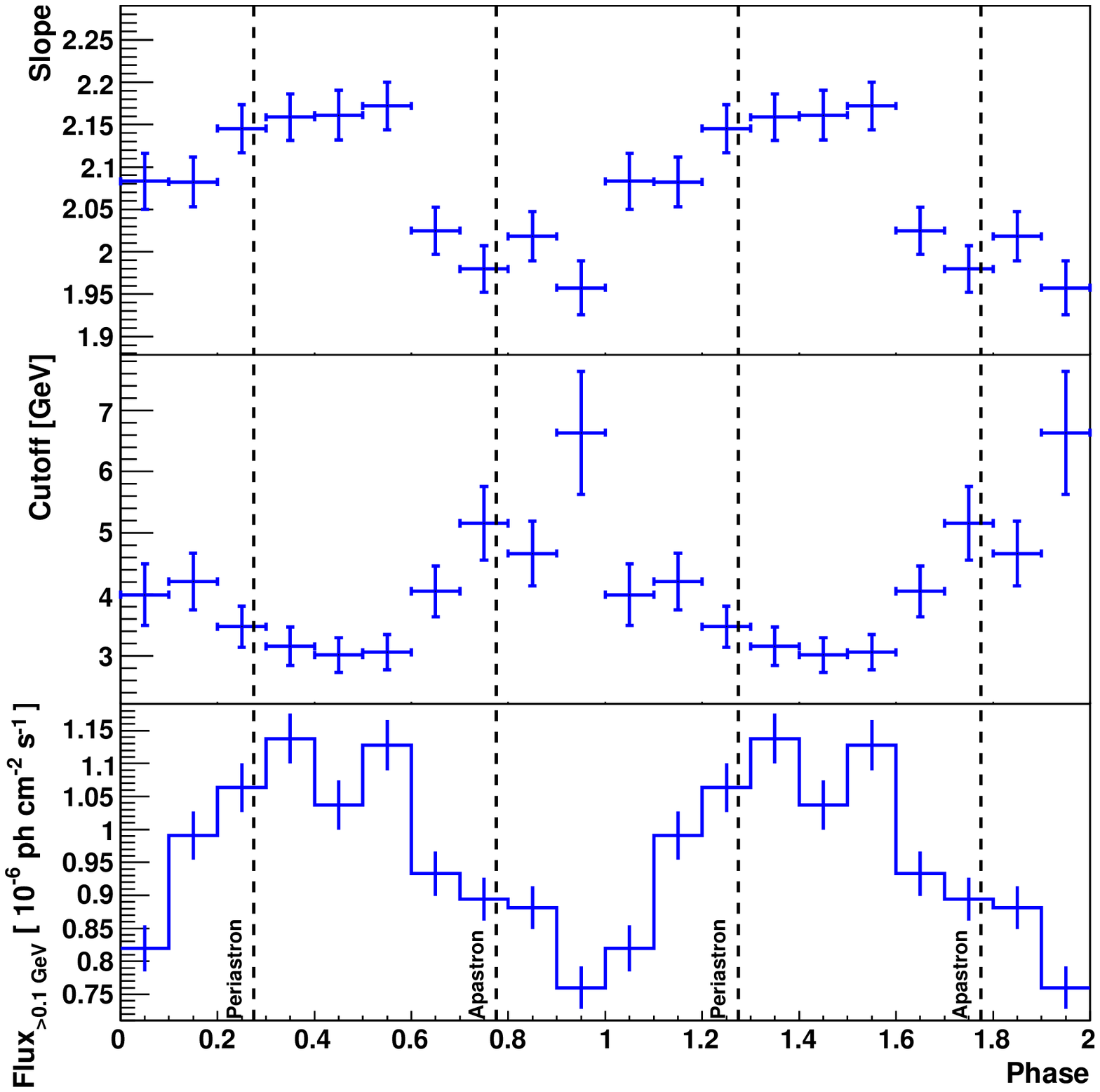}
\caption{\label{fig:lsi_modulation}The modulation of the spectral parameters of \lsi\ on the total 30 months lightcurve. For details see Section~\ref{phase}} 
    \end{center}
\end{figure}

\subsection{Lightcurve}\label{change}

In Figure~\ref{fig:orbital_lc_time_LSI} the lightcurve for
\lsi\ over 30 months is shown using orbital time bins. The black dashed line 
represents the point in time when a flux change occurred for \lsi; this
ocurred in March 2009. In 
Figure~\ref{fig:orbital_lc_time_LSI_timing} we present the power spectrum of \lsi\ 
derived from the total weighted photon lightcurve. The power spectrum clearly detects the orbital flux modulation
with a period of 26.71 $\pm$ 0.05 days. This is consistent with the known orbital period of \lsi\ \citep{2002ApJ...575..427G}. The lightcurve  clearly shows long-term variability.
We searched for changes in the orbital modulation of \lsi\ by dividing
the aperture photometry lightcurves into 6-month segments and calculating the power spectrum of each
segment, as shown in the left panel of Figure~\ref{fig:lsi_varying_power}. The
amplitude of the orbital modulation is estimated by fitting a sine wave fixed
to the orbital period to each of the lightcurve segments; the results, as
seen in the right panel of Figure~\ref{fig:lsi_varying_power}, clearly show a
decreasing trend in the orbital modulation with time.

\lsi\ is one of the brightest sources in the $\gamma$-ray sky and towers
above all other emitters in its neighborhood. This allows us to compute a
lightcurve with an orbital binning (26.496 days per bin) which is shown in Figure
\ref{fig:orbital_lc_time_LSI}. Even by eye it is clear that the source is highly variable on orbital
time scales and longer.  The longer term trends are evident by looking at a
plot of the 3-orbit rolling average (black line in Fig. \ref{fig:orbital_lc_time_LSI}). During the first eight orbits the flux
decreases by a factor of $\sim$2. Then, in March 2009, the flux appears to increase
over the course of several orbits; we take the transition point of this increase to be
MJD 54900.

The flux increases significantly by 33$\pm$4\%, rising from a baseline of
$(0.75 \pm 0.03_{\mathrm{stat}} \pm 0.07_{\mathrm{syst}}) \times 10^{-6} \,
\mathrm{ph\, cm^{-2}\, s^{-1}}$ obtained from the first 8
months of data to $(1.00 \pm 0.01_{\mathrm{stat}} \pm
0.07_{\mathrm{syst}}) \times 10^{-6} \, \mathrm{ph\, cm^{-2}\, s^{-1}}$ which is the
average flux of the remaining 1.7 years of the data. Comparing the
flux levels averaged over the same time span, 8 months before and 8 months
after the flux change, we obtain a $\sim$40\% increase. After this flux change the
flux decreases again slowly over the remaining 1.7 years.  The complexities of
the short timescale, orbit-to-orbit variability make it impossible to
characterise the exact properties of the transition from the `lower' to
`higher' flux states.  The transition likely took place over several orbits,
however, for simplicity throughout the remainder of this analysis we use a
transition time of MJD 54900.
 
We graphically
show the flux change in Figure~\ref{fig:before_after}, by plotting the folded
lightcurves before and after the transition in March 2009.
The data points are folded on the \citet{2002ApJ...575..427G} period,
with zero phase at MJD 43,366.775. Before the
transition, the modulation was clearly seen and is compatible with the already
published phasogram, whereas afterwards, the amplitude of the modulation
diminishes. We quantify this behavior by measuring the flux fraction below.
Note that the datasets corresponding to the reported results
\citep{2009ApJ...701L.123A} and what we here referred to as {\it before the
flux change}
span almost exactly the same time range, with the consequence of our current
analysis essentially reproducing that previously published. The time span
covered by our earlier publication coincidentally finished just prior to the
onset of the flux change. The spectra derived before and after this flux change
are shown in Figure~\ref{fig:before_after}, where the increase in flux is
also obviously visible.

\begin{figure*}[ht]
    \begin{center}
 \includegraphics*[width=0.49\textwidth,angle=0,clip]{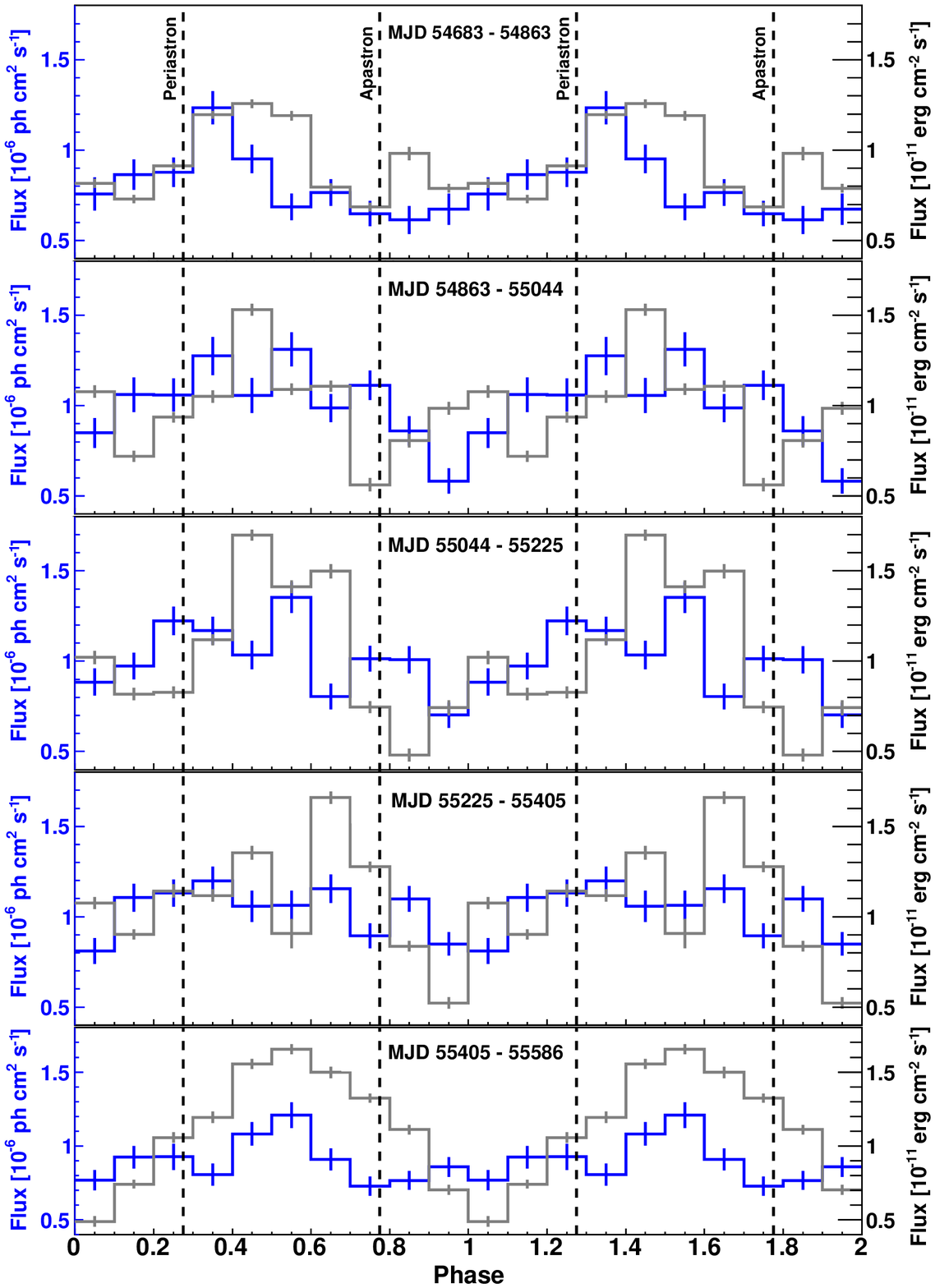}
    \includegraphics*[width=0.49\textwidth,angle=0,clip]{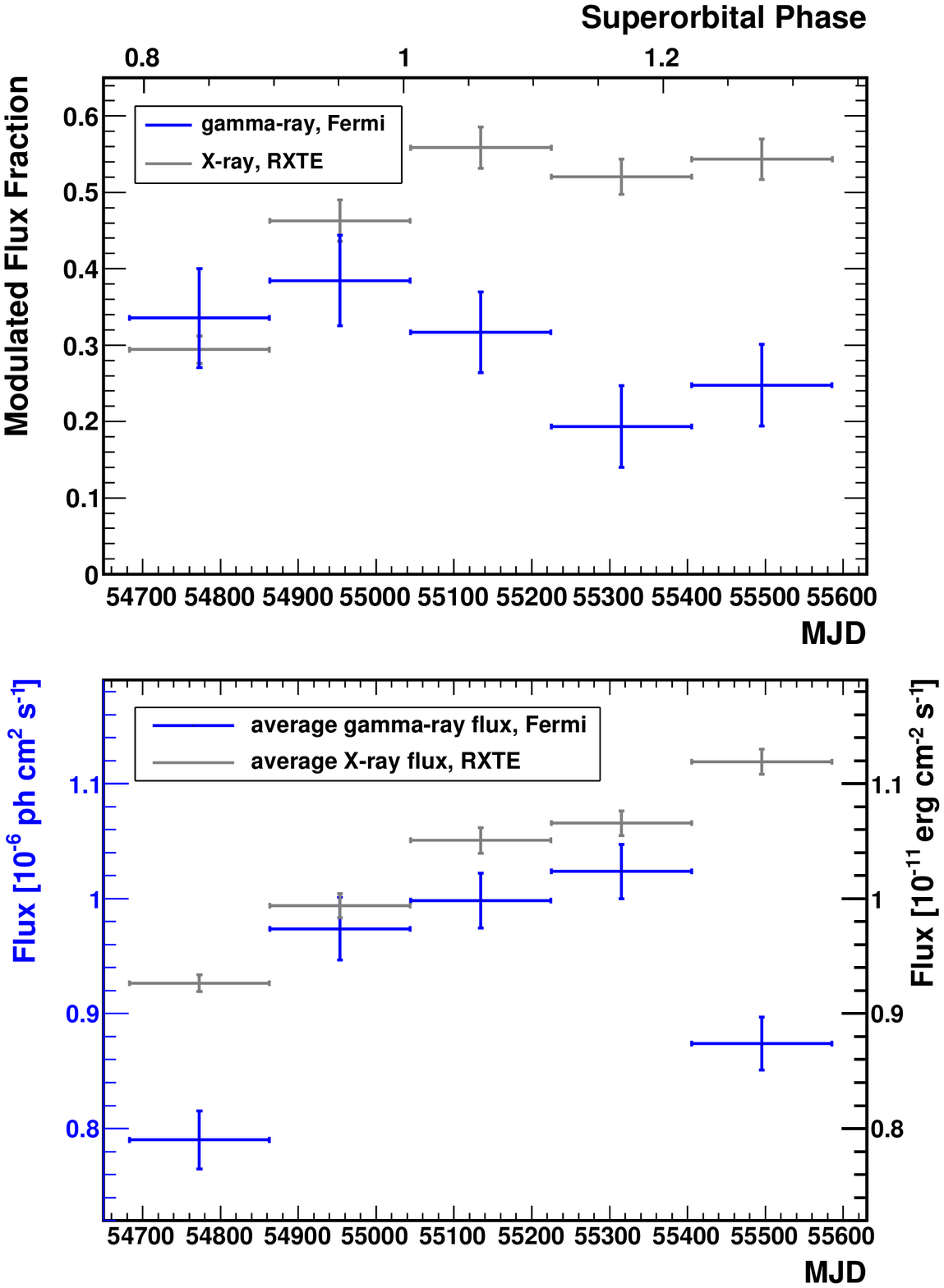}
\caption{\label{fig:multi}Comparison between the $\gamma$-ray (blue) and the X-ray (gray) data of \lsi\ .  \textit{Left: } For each of the 5 separate 6-months periods, the 3-10\,keV (gray) and the 100\,MeV-300\,GeV (blue) folded lightcurves are shown. \textit{Right: } 6-months modulated flux fraction data in both energy bands, with equal color coding.} 
    \end{center}
\end{figure*}

\begin{figure*}[ht]
    \begin{center}
    \includegraphics*[width=0.79\textwidth,angle=0,clip]{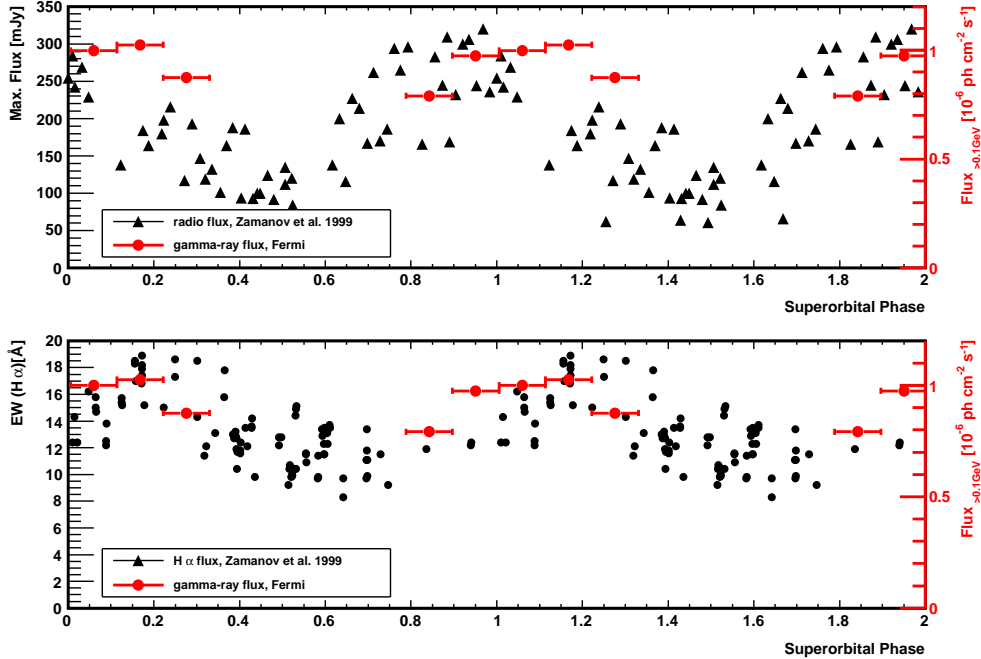}
\caption{\label{radio-comp}Comparison of non-contemporaneous LAT-obtained $\gamma$-ray with radio data of \lsi\ folded with the radio super orbital period.  \textit{Top: } With respect to the radio data compiled by Gregory (1999), mostly at 8.3 GHz.
 \textit{Bottom: } With respect to the EW of the H$\alpha$ line obtained by Zamanov et al. (1999).} 
    \end{center}
\end{figure*}

\begin{figure*}[th]
    \begin{center}
    \includegraphics*[width=0.79\linewidth,angle=0,clip]{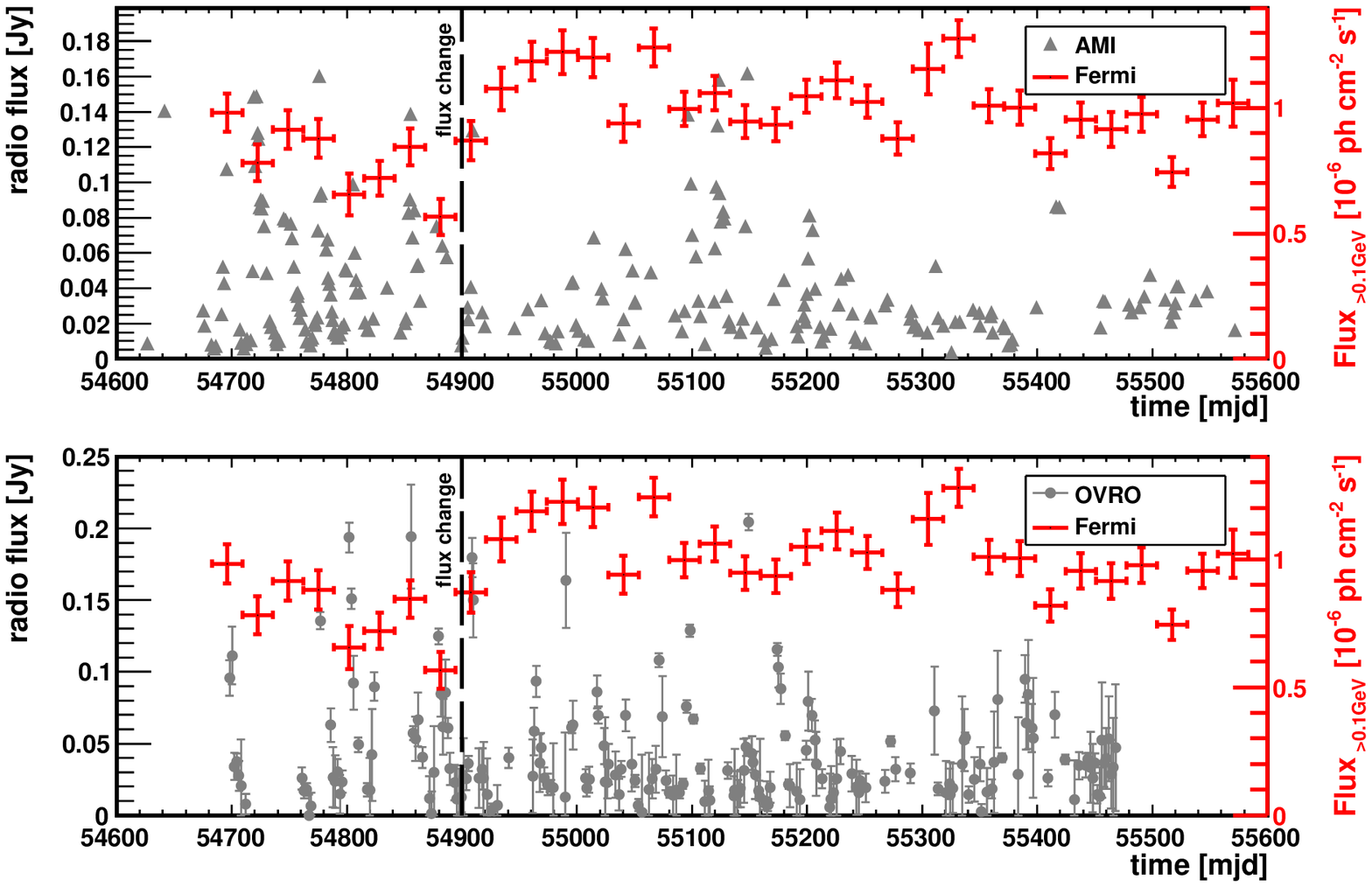}
\caption{\label{fig:whole_data_lc_lsi} Lightcurve of LS I +61$^\circ$303. Each point represent one orbit of 26.496 days. The black dashed line marks the moment of the flux change in March 2009, together with the simultaneously taken radio data by the AMI and OVRO instruments (both operating at 15\,GHz) are shown in gray. } 
    \end{center}
\end{figure*}

\subsection{Phase resolved spectral analysis}\label{phase} 

The statistics of the current dataset allow us to divide the orbit in different phase ranges and to compute the corresponding spectra for different phase bins.
We have divided the orbit into INFC and SUPC phase ranges in two different ways.
First, we have split  the \lsi\ orbit in two halves based on its geometry, as visualized, for instance in Aragona et al. (2009). The SUPC phase range is defined as from phase 0.63 to phase 0.13; INFC is defined correspondingly, as the remaining half.
We also adopted another way to separate between INFC and SUPC based on the 
angle between the compact object, the star, and the observer. Therefore,
the orbit is not divided in two halves in this way, but in one piece when the compact
object is in front of the star (corresponding to INFC): 0.244 -- 0.507; and in another phase range
with the same duration, centering at the exact SUPC phase: 0.981 -- 0.244.
As can be seen in Figure~\ref{fig:orbital_cuts}, the spectra obtained with the different cuts do not differ significantly and all spectral parameters are compatible within their corresponding errors (see Table~\ref{table:fit}). We find that the flux difference between INFC and SUPC is of order of 20\%.
For these different datasets, representing only portions of the orbit, 
we modeled the source with a pure power law and with a power law with an exponential cutoff.
The $\Delta TS$ value comparing both fits for the angular based cut is 108 (INFC), which means that the probability of incorrectly rejecting a power law with respect to an an exponential cutoff is $2.8\times10^{-25}$ ($10\sigma$).
For SUPC the $\Delta TS$ value is 97, which leads to a probability of $6.9\times10^{-12}$ ($6.8\sigma$) to wrongly reject the cutoff power law.
Hence, the exponential cutoff is preferred over the pure power law also for two parts of the orbit, namely INFC or SUPC.

We have also divided
the orbit into phase ranges corresponding to periastron (half of the orbit around phase 0.275)
and apastron (the other half, around phase 0.725), based just on the distance between the compact object and the star. In this case, neither a significant difference between the flux values for the two phase bins nor a difference in the spectral shape is visible. These results can also be seen in Figure~\ref{fig:orbital_cuts}. This is probably the result of dividing the orbit into phase ranges which contain both bin phases corresponding to INFC and SUPC.

We also studied the spectral behavior of the source in phase bins of 0.1.
For this study we modeled \lsi\ with a power law with cutoff for each phase-bin individually. The spectral parameters obtained are shown
in Figure~\ref{fig:lsi_modulation}.
Note that we fixed in the model the index at 2.07 to study the orbital behavior of the cutoff energy and we fixed the cutoff energy at 3.9\,GeV
to study the behavior of the index, since both parameters are correlated. These values are, respectively, the results arising from 
the fit over the whole dataset.
A clear orbital modulation of the flux and spectral shape are seen.
Through the periastron passage the spectrum gets softer and the flux is maximum, whereas around apastron the spectrum becomes harder and the flux reaches its minimum.

\subsection{Spectral fitting}\label{different_fits}

Both the orbitally-averaged spectrum and the spectra of the several datasets mentioned were also fit 
with a broken power law, in addition to the pure and exponentially cutoff
power law.  All the $TS$ values for the different fits are listed in
Table~\ref{table:TS}. It is evident that the power law with an exponential
cutoff or the broken power law always
describe the spectral shape better than a pure power law does.
To be precise, when
comparing the exponentially cutoff with the pure power law, the $\Delta
TS$~values span the range from 90 to 520; with the former being always statistically
preferred. 
Fitting a broken power law gives almost the same results. The $\Delta
TS$~values span the range from 77 to 494 and we find break energies in the range of 0.4 to 1.7\,GeV.
All of the fits to the various data sets with an exponentially cutoff power law have slightly better $TS$ values with fewer degrees of freedom than a broken power law suggesting that the former model is a better description of the data than the latter one.
The exponentially cutoff power law describes nicely the curvature of the
spectrum especially at low energies. 
At higher energies, 
above the spectral break, the broken power law fits all the spectral
points, even the highest one which might be considered possibly part of a second
component. 
Statistically we cannot distinguish which of these two fitting
models describes best the data as a whole, but only that both of them are
preferred over a pure power law.

\subsection{The multi-wavelength context}

\subsubsection{X-rays}

\lsi\ has also been monitored with the {\it RXTE}--Proportional Counter Array (PCA) and folded lightcurves 
were produced using the same ephemeris as described in Section~\ref{change}. In the left panel of Figure
\ref{fig:multi}, we show a direct comparison between the phasograms in X-ray
and in $\gamma$-rays, with simultaneously taken data. We divide the whole LAT
dataset into five periods of six months each and compare them with correspondingly
obtained PCA X-ray data. The division in periods of 6-months is justified in
order to have enough statistics in $\gamma$-rays for each individual time bin,
and such that orbit-to-orbit X-ray variability does not dominate the flux
fraction changes.

For {\it RXTE}-PCA, we have used the ``Standard 2'' mode for spectral analysis.
Data reduction was performed using \textsc{HEASoft 6.9}. 
We select time intervals where the source elevation above Earth limb is $>10^{\circ}$ and the pointing offset is $<0.02^{\circ}$.
PCA background lightcurves and spectra were generated using the \textsc{FTOOLS} task {\tt pcabackest}. {\tt pcarsp} was used to generate PCA response matrices for spectra.
The background file used in analysis of PCA data is
the most recent available from the HEASARC website for faint sources,
and detector breakdown events have been removed\footnote{The background file is pca\_bkgd\_cmfaintl7\_eMv20051128.mdl and see the website: http://heasarc.gsfc.nasa.gov/docs/xte/recipes/pca\_breakdown.html for more information on the breakdowns. The data have been barycentered using the FTOOLS routine faxbary using the JPL DE405 solar system ephemeris.}. A power law shape, with absorbing hydrogen column density fixed at $0.75\times10^{22}$ cm$^{-2}$ \citep{2005A&A...440..775K,2009ApJ...693.1621S}, was used to fit the Standard 2 data, with the following function $
N(E) = K e^{-N_H\,\sigma(E)} ( {E}/{{\rm keV}})^{-\alpha},$ where $K$ is a normalization at 1 keV, $\sigma$ is the photoelectric cross section and $\alpha$ is the photon index. 

It is apparent that the X-ray
modulation is always visible in each of these five panels, albeit with variable amplitude of flux modulation. 
We define the flux fraction as $(c_{max} - c_{min} )/(c_{max} + c_{min})$, where $c_{max}$ and $c_{min}$ are the maximum and minimum {flux in the 3--10 keV} found in the orbital profile analyzed (after background subtraction).
The modulation becomes even stronger over time and stays stable over the last 3 half year bins. Instead, at GeV energies,  
the LAT data indicate that the modulated fraction fades away until the variability along the orbit is barely visible in the last 6 months of our data, which is consistent with Figure \ref{fig:lsi_varying_power}.
The flux fraction is plotted in the right-top panel of Figure \ref{fig:multi}. The difference between the behavior of the X-ray and the
$\gamma$-ray emission is clearly visible. In the right-bottom panel of Figure \ref{fig:multi} we show the average $\gamma$- and X-ray fluxes fitted in each of the periods considered. 
We checked for a possible appearance of the super-orbital periodicity 1667$\pm$8 days, taking zero phase at MJD 43366.275 and the shape of the outburst peak flux modulation estimated by \citet{2002ApJ...575..427G} in radio by considering the flux evolution and the direct count rate, using the full dataset of the PCA observations. We find no evidence of the super-orbital period in X-ray or $\gamma$-rays, which is not surprising given  the short integration time in comparison with the super-orbital period duration.

\subsubsection{Radio and optical}

We have also compared the LAT-obtained data with radio and H$\alpha$
observations.  For the latter, we first take into account the results contained
in the long-term coverage presented by \citet{1999A&A...351..543Z}. These
authors have already shown that the equivalent width (EW) and the peak
separation of the H$\alpha$ emission line appear to vary with the super-orbital
radio period of $\sim 1600$ days, this being likely the result of cyclical
variations in the mass-loss rate of the Be companion and/or of density
variability in the circumstellar disk.  \citet{1999A&A...351..543Z} proposed
that the variability in the EW(H$\alpha$) can be explained by a cyclical change
in the mass-loss rate of about 25\% over its average value. This mechanism
would imply changes in the density of the circumstellar disc in the same range.
The fact that the $\gamma$-ray lightcurve would have a similar behavior to that found for
the EW(H$\alpha$) would imply that the former is related to the circumstellar
disc surrounding the Be star.  This would naturally be the case if the $\gamma$-rays
are produced in an inter-wind shock formed  by the collision of outflows from
the compact object and the star itself.
\citet{2006A&A...456..801D} and \citet{2009ApJ...693.1462S} discussed how the many uncertainties present in our knowledge of the \lsi\ system influence the models for producing $\gamma$-ray emission. One of the important parameters is the density and size (and the possible truncation) of the circumstellar disc. The latter influence the position of the inter-wind shock, and the time intervals along the orbit in which the compact object outflow may be balanced by the equatorial wind feeding it, and/or when the compact object is directly within the circumstellar disc itself. 
For recent measurements on very long-term optical variability of Be High Mass X-ray Binaries in the Small Magellanic Cloud see \citet{2011MNRAS.413.1600R}.

We plot in Figure \ref{radio-comp} a comparison of the LAT data, folded on the super-orbital radio period, with the radio data compiled  by Gregory (1999) -- most of it obtained at 8.3 GHz, using the Green Bank Interferometer --
and the EW of the H$\alpha$ line obtained by \citet{1999A&A...351..543Z}. 
Care should be exercised when comparing our LAT data with the radio data in \citet{1999A&A...351..543Z}, since the latter authors used the ephemeris given by \citet{1999ApJ...520..361G}, for which the combination of the phase and peak flux density yielded a best-modulation period of 1584 days. This super-orbital period was later revised by \citet{2002ApJ...575..427G} to the current value of 1667 days. This latter value is the super-orbital period that we have used to fold the X-ray and LAT data, and we use 
it also here for the radio and H$\alpha$ emission. 
The comparison is clearly non-contemporaneous, and this may induce problems in the interpretation of a source like \lsi, presenting different variability phenomenology.
We do not see any clear correlation or anti-correlation with the radio and H$\alpha$ fluxes which may be hidden by the scarcity of GeV data. 

We have also compared the Fermi data with simultaneous radio data
from the Owens Valley  Radio Observatory (OVRO) and the Arcminute Microkelvin Imager (AMI) array (Cambridge, UK). 
Since the launch of Fermi in June 2008, the OVRO 40\,m single-dish telescope,
that is located in  California (USA), has conducted a regular monitoring program of Galactic binaries (e.g. Cyg X$-$3,
 \citet{2009Sci...326.1512F}).  The OVRO flux densities are measured in a single 3\,GHz wide band centered on 15\,GHz.
A complete  description of the OVRO 40\,m telescope and calibration strategy can be found in \citet{2011ApJS..194...29R}.
Furthermore, we  also use complementary observations  provided by  AMI, consisting of a set of eight 13\,m antennas with a
maximum baseline of $\sim$ 120\,m. AMI observations are conducted with a 6\,GHz bandwidth receivers
also centered  at 15\,GHz. See \citet{2008MNRAS.391.1545Z} for more details on the AMI interferometer, that is mostly used
for study of the cosmic microwave background.
By folding these radio data no super-orbital modulation could be seen, which
could be due to the poor coverage of a whole super-orbit. A direct comparison of the
GeV and the radio data is shown in Figure~\ref{fig:whole_data_lc_lsi} where the
flux over time is plotted. No correlation of the data points can be seen either.
In summary, we did not find any correlation between the GeV and the radio band, either in 
archived nor in simultaneous data.


\section{Concluding remarks}

After analyzing a dataset comprising 2.5-years of \fermi-LAT observations of  the two binaries \ls\ and \lsi, we note several changes with respect to the initial reports of \citet{2009ApJ...701L.123A,2009ApJ...706L..56A}. These were produced either because the accumulation of a longer observation time allowed us to make distinctions that were earlier impossible (valid for both sources), or because  the behavior of the source changed (valid for \lsi).
On one hand, 
the statistics are now sufficient to divide the dataset of both sources in
INFC and SUPC and to show that a power law with a cutoff describes the spectra obtained in both conjunctions better than a pure power law. The cutoff is similar to that found in the many other GeV pulsars discovered by \fermi-LAT.
However, we have found that 
both \ls\ and \lsi\ show an excess of the high energy GeV emission beyond what
is expected from an exponentially cutoff power law.
While the high energy data are significantly in excess of the exponentially cutoff power law there are insufficient statistics at these high energies to model the excess with an additional spectral component.

The process(es) generating such a component in the case of \ls\ and \lsi\ is unclear and may even be different in the two sources.
However, such a second component would present a possible connection between the GeV and TeV spectra in both sources. Collecting more data and therefore more statistics will allow to prove or discard it in the future. The lack of datapoints at high energies, also affects, particularly in the case of \lsi, the distinction between an exponentially cut and a broken power law. Currently, both are certainly preferred over a pure power law, but differences in the significances provided between the former are minor.

We have noted that whereas \ls\ shows stable emission over time and also a stable orbital modulation,
\lsi\ shows a change in flux in March 2009. Afterwards the orbital modulation decreases (see the bottom-most left panels of Figure \ref{fig:multi})
and the orbital period could not be detected in the GeV data.
\lsi\ has also presented a complex, concurrent behavior at higher energies. At TeV, for approximately the last two years,
it seems to have been in a low state in comparison with the flux level that led to its discovery \citep{lowMAGIC2011,2011ApJ...738....3A}. Additionally, it was detected once  --after 4.2\,hrs of observations-- 
near the periastron, where the system was never seen at TeV energies before \citep{2011ApJ...738....3A}. Both of these aspects of the \lsi\ phenomenology --as well as the phase location of the TeV maxima in general--
would require modifications of 
simple inverse Compton models in the scenarios usually put forward for the source. 
The idea of a magnetar compact object in \lsi\ is suggested by \cite{Torres2011} to explain the changing TeV behavior of this object and could potentially describe the diminishing orbital modulation now observed at GeV fluxes.  However, as explained by \cite{Torres2011}, detailed numerical simulations are required to verify if this scenario can accurately reproduce the observed GeV and TeV emission.

The lower-energy, multi-wavelength picture of \lsi\ is still unclear with the orbital modulation in GeV (X-ray) fading (increasing) with time. Because the GeV data do not cover a whole super-orbital period, a conclusion about the possible relationship of the GeV emission with the super-orbital behavior of \lsi\ cannot be drawn yet. Continued monitoring of the source with the \fermi-LAT will allow the study such long-term cycles, if any.
Zamanov et al. (1999) concluded that if the $\sim$4\,yr modulation in radio and H$\alpha$ is due to changes in the circumstellar disk density, it could be detected in X- or $\gamma$-rays as well, since the high energy emission of a putative neutron star depends on the density of the surrounding matter with which it interacts. Indeed, this modulation was recently revealed in X-rays \citep{LiSuperOrbit}.

The nature of these systems is thus still unclear and multi-wavelength observations, including \fermi-LAT monitoring, as well as further theoretical studies are expected to bring new insights.

\mbox{}

{\textit{Acknowledgements:}}
The Fermi LAT Collaboration acknowledges generous ongoing support from a number of
agencies and institutes that have supported both the development and the operation of the LAT as
well as scientific data analysis. These include the National Aeronautics and Space Administration
and the Department of Energy in the United States, the Commissariat \`a l'Energie Atomique and
the Centre National de la Recherche Scientifique / Institut National de Physique Nucl\'eaire et de
Physique des Particules in France, the Agenzia Spaziale Italiana and the Istituto Nazionale di Fisica
Nucleare in Italy, the Ministry of Education, Culture, Sports, Science and Technology (MEXT),
High Energy Accelerator Research Organization (KEK) and Japan Aerospace Exploration Agency
(JAXA) in Japan, and the K. A. Wallenberg Foundation, the Swedish Research Council and the
Swedish National Space Board in Sweden.
Additional support for science analysis during the operations phase is gratefully acknowledged
from the Istituto Nazionale di Astrofisica in Italy and the
Centre National d'\'Etudes Spatiales in France.
This work has been additionally supported by the Spanish CSIC and MICINN and the Generalitat de Catalunya, through grants AYA2009-07391 and SGR2009-811, as
well as the Formosa Program TW2010005. SZ acknowledges supports from  National Natural Science Foundation of
China (via NSFC-10325313, 10521001, 10733010, 10821061, 11073021 and 11133002),
and 973 program 2009CB824800. GD acknowledges support from the European Community via contract ERC-StG-200911.
ABH acknowledges funding via an EU Marie Curie International Outgoing Fellowship under contract no. 2010-275861.
The AMI arrays are supported by STFC and the University of Cambridge.\\

\bibliographystyle{apj}

\end{document}